%% file: articleboltz_gr.tex
\documentclass{iopart}
\usepackage{graphicx,epsfig}
\expandafter\let\csname equation*\endcsname\relax
\expandafter\let\csname endequation*\endcsname\relax
\usepackage{amsmath}
\usepackage{bm}
\usepackage{epsfig}
\usepackage{color}
\def\roughly#1{\mathrel{\raise.3ex\hbox{$#1$\kern-.75em%
\lower1ex\hbox{$\sim$}}}}

\def\d{\mathrm{d}}

\makeatletter
\renewcommand*\env@matrix[1][\arraystretch]{%
  \edef\arraystretch{#1}%
  \hskip -\arraycolsep
  \let\@ifnextchar\new@ifnextchar
  \array{*\c@MaxMatrixCols c}}
\makeatother

\begin{document}

\title[GR Neutrino Transport]{General relativistic neutrino transport
  using spectral methods} 

\author{Bruno Peres, Andrew Jason Penner, J\'er\^ome Novak and Silvano Bonazzola}
\address{Laboratoire Univers et Th\'eories (LUTH), Observatoire de
  Paris/CNRS/Universit\'e Paris Diderot, 5 place Jules Janssen, 92195
  Meudon, France} 

\date{\today}


\begin{abstract}
  We present a new code, {\tt{Lorene's Ghost}} (for Lorene's
  gravitational handling of spectral transport) developed to treat the
  problem of neutrino transport in supernovae with the use of spectral
  methods. First, we derive the expression for the nonrelativistic
  Liouville operator in doubly spherical coordinates
  $(r,\,\theta,\,\phi,\,\epsilon,\,\Theta,\, \Phi)$, and further its
  general relativistic counterpart. We use the $3 + 1$ formalism with
  the conformally flat approximation for the spatial metric, to
  express the Liouville operator in the Eulerian frame. Our
  formulation does not use any approximations when dealing with the
  angular arguments $(\theta,\,\phi,\,\Theta,\,\Phi)$, and is fully
  energy-dependent. This approach is implemented in a spherical shell,
  using either Chebyshev polynomials or Fourier series as
  decomposition bases. It is here restricted to simplified collision
  terms (isoenergetic scattering) and to the case of a static
  fluid. We finish this paper by presenting test results using basic
  configurations, including general relativistic ones in the
  Schwarzschild metric, in order to demonstrate the convergence
  properties, the conservation of particle number and correct
  treatment of some general-relativistic effects of our code. The use
  of spectral methods enables to run our test cases in a
  six-dimensional setting on a single processor.
\end{abstract}

\section{Introduction}

Supernova numerical simulations is a rapidly growing field of
research. While the collapse, bounce, and prompt shock propagation
seem to be well reproduced by one-dimensional (1D) simulations
(\textit{e.g.}~\cite{lieb_04, ramp_02}), the late phases are clearly
multi-dimensional \cite{jank_12b, suwa_10, taki_12, dole_12,
  muel_12b}. Capturing the complete physical description of a
supernova explosion would require the combination of 3D hydrodynamics,
an accurate equation of state (EoS) as well as a six-dimensional (6D)
Boltzmann solver to handle the neutrino evolution. Unfortunately, such
a numerical simulation is not possible yet, because it would take a
prohibitive amount of computational time (see also \cite{kota_12}). In
such a simulation, the neutrino treatment would be the most demanding
computation. Within this context, general relativity (GR) has shown to
be a very important ingredient (see \textit{e.g.}~\cite{muel_12b,
  lent_12b}) and it is therefore necessary to devise
general-relativistic formulations and codes for neutrino numerical
simulations.

Radiative transfer is a widely studied field. When particle interactions are
sufficiently numerous to drive the system into local thermal equilibrium we may
treat the particles via their bulk properties, and thus as a fluid. However,
whenever interactions are not sufficient to drive the considered particle to
such an equilibrium, radiative transfer is used as the physical model. In the
supernova neutrino transport problem, this means solving a Boltzmann
equation\footnote{Some alternative approaches exist, such as a Monte--Carlo
  treatment of the radiative particle, as discussed in
  \cite{abdi_12}.}. Because the neutrinos opacity ranges continuously from
completely transparent up to completely opaque, and because the shock is known
to stall in the semi-transparent regime, having an accurate treatment of the
semi-transparent regime is essential. Furthermore, due to the demand for
highly detailed neutrino transport in multi-dimensional supernovae
simulations, on a physics point of view the treatment of the Boltzmann
equation should be done with as few simplifications as possible.

Due to the high computational cost of the full Boltzmann treatment,
approximations are often used, such as a light bulb approximation
or a leakage scheme. These approximations are encountered, for
example, when studying black hole formation (\cite{ocon_11, ugli_12}),
or when studying explosions in 3D in a reasonable amount of time
(\cite{dole_12, ott_12b, hank_11, nord_10}). One of the aims of the
modelers is to minimize the approximations used in these studies.

An accurate treatment of the Boltzmann equation in spherical symmetry
was presented in \cite{lieb_04}. Several approximate treatments
coupled to multi-dimensional hydrodynamics were later proposed, and we
make a brief list hereafter. The diffusion approximation is a method
used to approximately solve the Boltzmann equation in the supernova
context (see, \textit{e.g.}~\cite{brue_12}). It is well known that
the diffusion model, relying on parabolic type partial differential
equations, can generate superluminal propagation
velocities\footnote{ We refer the reader to \textit{e.g.}~\cite{bona_11} for a
  possible method to solve this problem by use of the telegraph equation.}. In
\cite{shib_11}, the authors also note that the diffusion approximation
could potentially violate the constraints in the Einstein
equations. The isotropic diffusion source approximation (IDSA) of
\cite{lieb_09} treats the neutrinos in the opaque regime and in the
transparent regime. An elegant interpolation between the two permits a
good approximation of the semi-transparent regime. Unfortunately, it
is not very well suited for an adaptation to general relativity, for
very similar reasons as the diffusion approximation.

In \cite{muel_10}, the authors implement a general relativistic
Boltzmann equation by means of a ray-by-ray ``plus'' approximation
with momentum closure. While these approximations are well-suited for
a general relativistic treatment, it is not obvious to which extent
the approximations used would compare to a Boltzmann treatment without
approximation. The ray-by-ray approximation treats a series of
spherically symmetric Boltzmann equations for different angles (the
$\theta$ direction in this paper). The lateral coupling (including the
neutrino coupling) is done using hydrodynamics. The momentum closure,
also proposed by \cite{shib_11} is an approximation which is even
more difficult to quantify. While the GR Boltzmann equation of
\cite{muel_10} has been used with 2D-axisymmetric hydrodynamics in
supernovae simulations (e.g. \cite{muel_12b}), recently, in
\cite{hank_13} the same group used the ray-by-ray ``plus'' in a 3D
supernova simulation, but without the general relativistic formalism.

A nonrelativistic Boltzmann equation with no approximation in the
angular dimensions has been addressed in \cite{livn_04, ott_08}, with
the use of $S_n$ methods (or discrete ordinates, see also
\cite{lieb_04}). Their approach neglects the fluid velocity, while
most others approximate it to $O(v)$ (an exception is the code
described in~\cite{lieb_04} working with Lagrangian coordinates). The
no-velocity-contribution issue has been investigated in detail in
\cite{lent_12a}. The authors found that completely neglecting
velocities is rather problematic in the study of stellar
core-collapse.

The rapidly growing literature in supernovae simulations points
towards a 3D hydrodynamics simulation. It has been done recently by
\cite{hank_13}, and by the authors of \cite{taki_12}, that implemented
a 3D ray-by-ray version of IDSA to tackle the neutrino transport. On
the other hand, several authors solve a radiative transfer equation
that is capable of handling 6D neutrino transport with full three
dimensional treatment in the coordinate space, without relying on an
approximation in the angular dimensions, either with the $S_n$
methods~\cite{sumi_12}, or with so-called $P_n$ methods, where the
angular part of the momentum space is decomposed onto a set of
spherical harmonics~\cite{radi_12}. Note that these formulations do
not include GR yet. Finally, \cite{shib_11} and \cite{card_12} showed
in great detail formalisms adapted to the resolution of the general
relativistic Boltzmann equation; and in \cite{kuro_12} the authors
employ the formalism developed in \cite{shib_11} with an approximate
neutrino treatment, including 3D simulations.

Here we present a general-relativistic version of the Boltzmann
equation, for the treatment of the neutrino transport in core-collapse
supernova simulations. Both in the relativistic and nonrelativistic
cases, the distribution function, $f$, giving information on the number
density of neutrinos in the phase space, depends on spacetime and
momentum coordinates. We show a double spherical system of such
coordinates and devise numerical methods that are adapted to such a
system. The left-hand side of the Boltzmann equation contains a
differential operator acting on $f$, namely the \emph{Liouville
  operator\/}. The right-hand side contains the \emph{collision
  terms\/} describing neutrinos reactions at the microphysical
level. We use static configurations in our test cases, because we do
not couple our neutrinos to a fluid, thus dropping the velocity terms
is fully justified.

As briefly discussed above, different authors have implemented a
treatment of the Boltzmann equation to describe the neutrino
transport. We, for the first time, propose to tackle it in GR using
spectral methods. This work closely follows the nonrelativistic work
by \cite{bona_11}. Spectral methods (\textit{e.g.}~\cite{gran_09} and
references therein) decompose a function using global basis functions,
say, Fourier or Chebyshev polynomials, instead of treating it locally
(as the finite difference approximation does). We take advantage of
the very rapid convergence expected from spectral methods when solving
practical applications presented in this paper. As a result, one can
use fewer points than with finite differences to get to the same
resolution, yielding a substantial decrease in computational time.

This paper is organized as follows. In Sec.~\ref{sec:nonrel} we link our work
to previous studies and define the Liouville operator in so-called doubly
spherical coordinates. In Sec.~\ref{sec:gr} we carefully derive a $3+1$
general relativistic Boltzmann equation with the Liouville operator in the
Eulerian frame, in a way that is particularly suited to be solved via spectral
methods, before exposing our results on test cases in
Sec.~\ref{sec:testresults}. Numerical methods are detailed in
\ref{app:implementation}.

In this paper we use a metric signature $(-, +, +, +)$ and geometrical
units in which $c = G = \hbar = 1$. Greek indices run from 0 to 3, while
Latin indices run from 1 to 3. We adopt the Einstein summation convention.

\section{Liouville operator in doubly spherical coordinates}
\label{sec:nonrel}

\subsection{Doubly spherical coordinates}
\label{ss:doubly_spher}

In stellar core-collapse simulations it is often convenient to assume
symmetries (spherical, \textit{e.g.}~\cite{lieb_04}, or axial,
\textit{e.g.}~\cite{muel_10}) in the problem, in order to decrease the
dimensionality and thus, the required computer power. The doubly
spherical system of coordinates (see \textit{e.g.}~\cite{pomr}) is
well-adapted to such approaches.
In general relativity, the time coordinate can be
defined within the $3 + 1$ formalism, as described \textit{e.g.} in~\cite{3plus1},
and the momentum four-vector must fulfill the mass-shell condition,
resulting in only three independent components. So, in both relativistic
and nonrelativistic cases, the distribution function
depends on time, three spatial coordinates $\left\{ x^i \right\}$
and three momentum coordinates $\left\{ p^i \right\}$.
\begin{figure}
  \centering
  \def\svgwidth{200pt}
    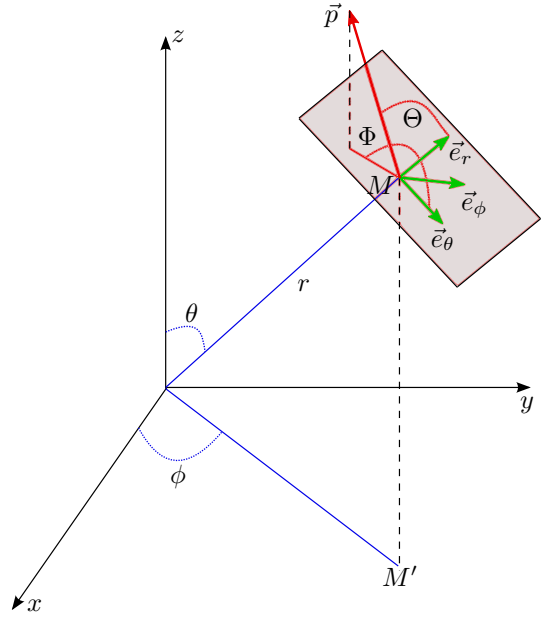
  \caption{Representation of the 6D double spherical coordinate
    basis.}
  \label{figure:doublespherical6d}
\end{figure}
At a given time, a point $\left( M, \vec{p} \, \right)$ in phase-space is
described as in Fig.~\ref{figure:doublespherical6d} in the following
way. Spatial coordinates are given in the usual spherical polar form:
one first defines the Cartesian triad $\left( \vec{e}_x, \vec{e}_y,
  \vec{e}_z \right)$\footnote{Here, all vectors $\vec{e}_{i}$ are
  unitary.}. Then $\overrightarrow{OM} = r\, \vec{e}_r$ defines
$r$ and the spherical basis vector $\vec{e}_r$. $\theta$ is the angle
between $\vec{e}_z$ and $\vec{e}_r$ and, with $M'$ being the
projection of $M$ onto the $\left( \vec{e}_x, \vec{e}_y \right)$
plane, $\phi$ is the angle between $\vec{e}_x$ and
$\overrightarrow{OM}'$.

Similarly, one defines $\vec{p} = \epsilon\,\vec{e}_\epsilon$, and
$\Theta$ is the angle between $\vec{e}_r$ and $\vec{e}_\epsilon$ (see
Fig.~\ref{figure:doublespherical6d}). Let $\vec{p}\,'$ be the projection
of $\vec{p}$ onto the $\left( \vec{e}_\theta, \vec{e}_\phi \right)$
plane; one then defines $\Phi$ as the angle
between $\vec{e}_\theta$ and $\vec{p}\,'$. Thus, the neutrino
distribution function depends on the 6 coordinates $\left(r, \theta,
  \phi, \epsilon, \Theta, \Phi \right)$ and time. From these
definitions and with the help of Fig.~\ref{figure:doublespherical6d},
we get
\begin{equation}
  \label{eq:pcartreal}
  \begin{pmatrix}
    p^r \\ p^\theta \\ p^\phi 
  \end{pmatrix} 
  = 
  \begin{pmatrix}
    \epsilon \cos \Theta \\ \epsilon \sin \Theta \cos \Phi \\
    \epsilon \sin \Theta \sin \Phi 
  \end{pmatrix}.
\end{equation}
Expressing $p^i$ in the Cartesian $\vec{e}_x, \vec{e}_y, \vec{e}_z$
basis would imply a second transformation from a spherical basis to a
Cartesian basis (done in~\ref{app:transfomatrix}), hence the name
``doubly spherical coordinates''.

Finally, let us introduce the momentum space Jacobian, which is needed
to express the Liouville operator in the Eulerian frame. The Jacobian
arises because we change the momentum derivative from the $\left(
  \vec{e}_r, \vec{e}_\theta, \vec{e}_\phi\right)$ basis
to the doubly spherical basis in momentum space $(\vec{e}_\epsilon, \vec{e}_\Theta,
\vec{e}_\Phi)$, see Eq.(\ref{eq:boltzefcfc}). The Jacobian is expressed as
\begin{equation}
  \label{eq:momspcjacobian}
  J^i\,\!_j = \dfrac{1}{\epsilon}
  \begin{pmatrix}[2]
    \epsilon \cos \Theta, & \epsilon \sin \Theta \cos \Phi, & \epsilon
    \sin \Theta \sin \Phi, \\
    -\sin \Theta, & \cos \Theta \cos \Phi, & \cos \Theta \sin \Phi, \\
    0, & -\dfrac{\sin \Phi}{\sin \Theta}, & \dfrac{\cos \Phi}{\sin \Theta},
  \end{pmatrix},
\end{equation}
(see \textit{e.g.}~\cite{ende_12c}). Here $j$ is the column index and $i$ is
the row index.

\subsection{Nonrelativistic Liouville operator without external
  forces}\label{ss:nonrel_liouville}

The nonrelativistic Boltzmann equation can be written as
\begin{equation}
  \label{eq:boltznonrel}
  \frac{\partial f}{\partial t} + \frac{1}{\epsilon}
  \frac{\d{}x^i}{\d{}t} \frac{\partial f}{\partial x^i} 
  + \frac{1}{\epsilon} \frac{\d{}p^i}{\d{}t}
  \frac{\partial f}{\partial p^i} = C[f],
\end{equation}
where $C[f]$ is the collision term. In classical theory,
one can identify the external forces represented by the term
$\d{}p^i/\d{}t$\footnote{In the case of Boltzmann equation, these are
  the infinite range forces.}. In doubly spherical coordinate, a
difficulty arises from the fact that the two momentum space angles
$\Theta,\,\Phi$ are, themselves, functions of the two real space
angles $\theta, \phi$ (see figure~\ref{figure:doublespherical6d} and
\cite{pomr, bona_11}).

From Eq.~(\ref{eq:pcartreal}) and the definitions of $p^i$ and $x^i$,
we write the spatial part of the Liouville operator,
\begin{equation}
  \label{eq:realspaceop}
  \frac{p^i}{\epsilon}\frac{\partial f}{\partial x^i} = \cos \Theta
  \frac{\partial f}{\partial r} + \frac{\sin \Theta \cos \Phi}{r}
  \frac{\partial f}{\partial 
    \theta} + \frac{\sin \Theta \sin \Phi}{r \sin \theta}
  \frac{\partial f}{\partial \phi}. 
\end{equation}
To write the complete Liouville operator without external forces, we
need to express the terms coming from the momentum-space part. They
arise from the fact that $p^i$ depends on $\theta$ and $\phi$.

We take the total temporal derivative of the momentum and explicitly
express the functional dependence on $\theta$, $\phi$
\begin{equation}
  \label{eq:chainrule}
  \left. \frac{\d{}p^i}{\d{}t} \right|_{\rm total} = \left. \frac{\d{}p^i}{\d{}t} \right| _{\theta, \phi}
  + \frac{\d\theta}{\d{}t} \frac{\partial p^i}{\partial \theta}
  + \frac{\d\phi}{\d{}t} \frac{\partial p^i}{\partial \phi}.
\end{equation}
We use Eq.~(\ref{eq:chainrule}) in Eq.~(\ref{eq:boltznonrel}), bearing
in mind that $\frac{\d{}p^i}{\d{}t}$ in Eq.~(\ref{eq:boltznonrel})
represents external forces and corresponds to
$\left. \frac{\d{}p^i}{\d{}t} \right| _{\theta, \phi}$ here. The other
terms in equation~(\ref{eq:chainrule}) are purely geometric, and come from the
fact that the angles $\Theta$ and $\Phi$ are defined with respect to axis that
are not fixed in space (as discussed in~\cite{pomr}). One can
then rewrite equation~(\ref{eq:boltznonrel}) taking this into account,
\begin{equation}
  \label{eq:boltznonrelwchrule}
  \frac{\partial f}{\partial t} + \frac{1}{\epsilon}
  \frac{\d{}x^i}{\d{}t} \frac{\partial f}{\partial x^i} 
  + \frac{1}{\epsilon}  \left( \left. \frac{\d{}p^i}{\d{}t} \right|
    _{\rm total} - \frac{\d\theta}{\d{}t} \frac{\partial p^i}{\partial
     \theta} - \frac{\d\phi}{\d{}t} \frac{\partial p^i}{\partial \phi}
  \right) \frac{\partial f}{\partial p^i} = C[f].
\end{equation}
Suppose that no external forces are present, then
$\left. \frac{\d{}p^i}{\d{}t} \right| _{\rm total} = 0$ but the
extra terms involving momentum remain. To compute these extra momentum
expressions, note that
\begin{equation}
  \frac{\d\theta}{\d{}t} = p^{\theta} \quad \textrm{and} \quad \frac{\d\phi}{\d{}t} = p^{\phi},
\end{equation}
so that we have to compute
\begin{equation}
  \label{eq:pthpphi}
  p^{\theta}\frac{\partial p^i}{\partial \theta}\frac{\partial
    f}{\partial p^i} \quad \textrm{and} \quad p^{\phi}\frac{\partial p^i}{\partial \phi}\frac{\partial f}{\partial p^i}.
\end{equation}

To compute expressions in Eq.~(\ref{eq:pthpphi}), we make use of the
transformation matrix $P = \dfrac{\partial \epsilon, \Theta,
  \Phi}{\partial r, \theta, \phi}$. After some algebra detailed in
\ref{app:transfomatrix} we obtain
\begin{equation}
  \label{eq:jac2}
  P = \begin{pmatrix}[2]
0, & 0, & 0, \\
0, & \cos \Phi, & \sin \Phi \sin \theta, \\
0, & - \dfrac{\sin \Phi \cos \Theta}{\sin \Theta}, & \cos \theta + \sin
\theta \dfrac{\cos \Phi \cos \Theta}{\sin \Theta},
\end{pmatrix}~.
\end{equation}

We thus get for the term involving~$\theta$
\begin{equation}
  \label{eq:termsupth}
  p^{\theta}\frac{\partial p^i}{\partial \theta}\frac{\partial
    f}{\partial p^i} = p^\theta \left[ \frac{\partial \Theta}{\partial
      \theta} \frac{\partial f}{\partial \Theta} +
    \frac{\partial \Phi}{\partial \theta}\frac{\partial f}{\partial
      \Phi} \right]
  = p^\theta \left[ \cos \Phi \frac{\partial
      f}{\partial \Theta} - \frac{\sin \Phi \cos \Theta}{\sin \Theta}
    \frac{\partial f}{\partial \Phi} \right].
\end{equation}
Analogously, for the term involving $\phi$, we get
\begin{equation}
  \label{termsupphi}
  p^{\phi}\frac{\partial p^i}{\partial \phi}\frac{\partial
    f}{\partial p^i} = p^\phi \left[ \frac{\partial \Theta}{\partial
      \phi} \frac{\partial f}{\partial \Theta} +
    \frac{\partial \Phi}{\partial \phi}\frac{\partial f}{\partial
      \Phi} \right] = p^\phi \left[ \sin \phi \sin \theta
    \frac{\partial f}{\partial \Theta} \right. 
     + \left. \left( \cos \theta + \sin
      \theta \frac{\cos \Phi \cos \Theta}{\sin \Theta} \right)
    \frac{\partial f} {\partial \Phi} \right]~.
\end{equation}

Combining equations~(\ref{eq:realspaceop}),
(\ref{eq:boltznonrelwchrule}), (\ref{eq:termsupth}) and
(\ref{termsupphi}), we get the nonrelativistic Liouville operator in
doubly spherical coordinates without external forces, $\mathcal{L}[f]$:

\begin{eqnarray}
  \mathcal{L}[f] &=& \frac{\partial f}{\partial t} + \cos \Theta
  \frac{\partial f}{\partial r} +
  \frac{\sin \Theta \cos \Phi}{r} \left[ \frac{\partial f}{\partial
      \theta} - \cos \Phi \frac{\partial f}{\partial \Theta} 
  +  \frac{\sin \Phi \cos \Theta}{\sin \Theta}
    \frac{\partial f}{\partial \Phi} \right] \nonumber\\ &+&
  \frac{\sin \Theta \sin 
    \Phi}{r \sin \theta} \left[ \frac{\partial f}{\partial \phi} - \sin \Phi \sin \theta \frac{\partial f}{\partial \Theta} \right.
  \nonumber\\ &+& \left. \left( \cos \theta + \sin
      \theta \frac{\cos \Phi \cos \Theta}{\sin \Theta} \right) \frac{\partial
      f}{\partial \Phi} \right].
\end{eqnarray}

After some algebraic simplification, the desired formula becomes,

\begin{eqnarray}
  \label{eq:fullspatialnonrel}
  \mathcal{L}[f] &=& \frac{\partial f}{\partial t} + \cos \Theta
  \frac{\partial f}{\partial r} + \frac{\sin \Theta \cos \Phi}{r} \frac{\partial f}{\partial
    \theta} +  \frac{\sin \Theta \sin \Phi}{r \sin \theta}
  \frac{\partial f}{\partial \phi} \nonumber\\ &&- \frac{\sin \Theta}{r} \frac{\partial
    f}{\partial \Theta} - \frac{\sin \Theta}{r} \sin \Phi \frac{\cos
    \theta}{\sin \theta} \frac{\partial f}{\partial \Phi}.
\end{eqnarray}

This expression agrees with those stated in \cite{pomr} and
\cite{bona_11}. Finally, note that Eq.(\ref{eq:fullspatialnonrel}) can be
derived with general relativistic formulation. Indeed,
Eq.(\ref{eq:boltzeffull}) divided by $\epsilon$ (for a question of different
notations and usage between the relativistic and non relativistic Boltzmann
equation only), which is the most general Boltzmann equation for
neutrino transport, reduces to Eq.(\ref{eq:fullspatialnonrel}) when
considering a vanishing velocity and a flat metric in spherical coordinates.

\subsection{General-relativistic equation}

The derivation of the covariant Boltzmann equation is not given here
and we refer the reader to \cite{stewart, card_05, debb_09, debb_09b},
which perform it in explicit detail. We present the general
expression of this equation below for convenience:

\begin{equation}
  \label{eq:b1}
  \epsilon \frac{\partial f}{\partial t} + p^i
  \frac{\partial f}{\partial x^i} - \Gamma^i\,\!_{\mu \nu}
  p^\mu p^\nu \frac{\partial f}{\partial p^i} = \mathcal{C}[f].
\end{equation}

Here $x^i$ is the position, $p^i$ the momentum, $\Gamma^i{}_{\mu \nu}$
the Christoffel symbols, and $\epsilon = p^0$. The left hand side of
the equation is the general relativistic Liouville operator, that we
note $L[f]$, and the right hand side is the collision operator. When
using this equation to solve the supernova neutrino transport problem,
we must take note of the different reference frames that are more
convenient to compute the different parts of the Boltzmann
equation. This step is explained in detail in section~\ref{sec:gr}. In
particular, $L[f]$ and $\mathcal{C}[f]$ (as well as the distribution
function $f$ itself) are scalars, so their expression does not depend
on the reference frame. We shall express $L[f]$ in the local inertial
frame (Eulerian frame, see Sec.~\ref{ss:frames}), while the collision
operator is expressed in the fluid rest frame (Lagrangian frame, see
Sec.~\ref{ss:frames}) and involves the neutrino 4-momentum as seen by
the Lagrangian observer in the fluid rest frame. We thus have the
mass-shell condition (expressed in Minkowski metric of the fluid rest
frame)
\begin{equation}
  \label{eq:mass_shell}
  p^\mu p_\mu = 0 = -\epsilon^2 + p^i p_i,
\end{equation}
and we have that $\epsilon$ still represents the norm of the 3-momentum
$p^i$. Note again that the result for $L[f]$ without gravity is the same as
the nonrelativistic one without external forces $\mathcal{L}[f]$
(Eq.~\ref{eq:fullspatialnonrel}).

\section{A $3+1$ formulation of the general relativistic Liouville
  operator}\label{sec:gr}

\subsection{Introduction}

In this section, we derive a $3 + 1$ form of the general relativistic
Liouville operator that is convenient to solve with spectral methods, because 

\begin{enumerate}
 \item It restricts the time derivatives to the distribution function,
   not on any of the dependent variables, 
 \item Its computation avoids the axial singularities that arise from
   terms which contain $\sin^{-1}(\theta)$ or $\sin^{-1}(\Theta)$ (see
   \ref{app:singularities}, also for the discussion of the
   singularity at the center). 
\end{enumerate}

We use Eq.~(\ref{eq:b1}) as the starting point for our derivation. This
expression is valid in the coordinate frame, with coordinate frame
momentum. Unfortunately, the collision operator, $\mathcal{C}[f]$, is
more easily computed in the fluid rest frame, or equivalently Lagrangian
frame. In this section, we carefully define these frames including the
Eulerian frame, and how one can transform an equation from one frame
to another. We chose to express the Boltzmann equation in the Eulerian
frame, which is more convenient when using spectral methods and using
a $3+1$ decomposition. This type of derivation has already been shown
in \cite{lind_66, card_05, ende_12c}.

Hereafter, to distinguish between different frames, we note fluid rest
frame variables with hatted indices ($\hat{\mu}$), coordinate frame
variables with tilde indices ($\tilde{\mu}$), and Eulerian frame
variables with unadorned indices ($\mu$). These frame conventions agree
with the literature (\cite{3plus1,muel_10}), although some ambiguity
does arise.

For convenience, we write the full Boltzmann equation with a
detailed Liouville operator involving Eulerian quantities (see
\ref{ss:frames} for the definition of the different frames). We detail
all these quantities in this section, and specialize our equation to a
static fluid and to a conformally flat spacetime (see
section~\ref{sec:cfc}). The full Boltzmann equation reads
\begin{equation}
  \label{eq:boltzeffull}
  \Lambda^\nu\,\!_{\hat{\mu}} p^{\hat{\mu}}
  M^{-1}\,\!\!_\nu\,\!^{\tilde{\rho}} \frac{\partial
    f}{\partial x^{\tilde{\rho}}} - \left(
    \Lambda^{-1}\,\!\!_\kappa\,\!^{\hat{n}} \Lambda^\zeta\,\!_{\hat{\mu}}
    M^{-1}\,\!\!_\zeta\,\!^{\tilde{\beta}} \frac{\partial
      \Lambda^\kappa\,\!_{\hat{\alpha}}}{\partial
      x^{\tilde{\beta}}} + \bar{\Gamma}^j\,\!_{\zeta \nu}
    \Lambda^\zeta\,\!_{\hat{\mu}} \Lambda^\nu\,\!_{\hat{\alpha}}
    \Lambda^{-1}\,\!\!_j\,\!^{\hat{n}} \right) p^{\hat{\mu}} p^{\hat{\alpha}}
  J^{\hat{k}}\,\!_{\hat{n}}
  \frac{\partial f}{\partial p^{\hat{k}}} =
  \mathcal{C}[f],
\end{equation}
where $\Lambda$ is a Lorentz boost (see Eq.~(\ref{eq:boost})), that
depends on the Eulerian 3-velocity of the fluid $v^i$. $M^{-1}$ is a
tetrad transformation matrix (see section~\ref{sect:tetrad}),
$p^{\hat{\mu}}$ is the neutrino 4-momentum seen by the Lagrangian
observer (see Eq.~(\ref{eq:boost})), $\bar{\Gamma}$ are the Ricci
rotation coefficients (see section~\ref{sec:riccicoeffs}), and $J$ is
the momentum space Jacobian (Eq.~(\ref{eq:momspcjacobian})). This
equation agrees with \textit{e.g.}~\cite{muel_10, card_05}.

\subsection{Different frames in general relativity}\label{ss:frames}

Below we discuss the coordinate frame, the Lagrangian frame, and the Eulerian frame.

The \emph{coordinate frame} (CF) is associated with the natural
basis. In numerical relativity, most of the time it coincides with the
numerical grid. For instance, in spherical geometry the basis vectors
$\vec{\partial}_t,\,\vec{\partial}_r,\,\vec{\partial}_\theta,\,\vec{\partial}_\phi$,
span the coordinate frame basis. This basis is fixed and the CF is not
associated with any physical observer, as $\vec{\partial}_t$ may
become spacelike. We choose spherical polar coordinates, namely
$r,\,\theta,\,\phi$; the components of the metric $\bm{g}$ in the CF
are denoted $g_{\tilde{\mu} \tilde{\nu}}$.

The \emph{Lagrangian frame} (LF), or fluid rest frame, is associated
with an observer comoving with the fluid. The associated observer
would therefore have the 4-velocity of the fluid. In the supernova
neutrino transport problem, the collision operator $\mathcal{C}[f]$ in
the Boltzmann equation involves quantities which are simpler to
compute in the LF. So, it is quite unavoidable to define the
components of the 4-momentum in the LF. It is the main reason why we
need these frame manipulations.

The \emph{Eulerian frame} (EF) is the locally inertial frame in
general relativity. It is the frame associated with the Eulerian
observer, who moves orthogonally to spacelike hypersurfaces of
constant time coordinate (see \textit{e.g.}~\cite{3plus1}). In the Eulerian
frame, locally, the four basis vectors
$\vec{e}_0,\,\vec{e}_1,\,\vec{e}_2,\,\vec{e}_3$ span an orthonormal
basis. In this basis, the metric is represented by a Minkowski matrix
\begin{equation}
  \label{eq:mink}
  \eta_{\mu \nu} = \bm{g} (\vec{e}_\mu, \vec{e}_\nu).
\end{equation}
Adopting a 3+1 representation, only $\vec{e}_0$ is a timelike vector,
and so $\vec{e}_1,\,\vec{e}_2,\,\vec{e}_3$ are spacelike. Transforming
from the CF to the EF (or vice-versa) is described in
Sect.~\ref{sect:tetrad} by means of a tetrad approach. The Eulerian
observer sits on spacelike hypersurfaces and sees events that are
causally linked to him. The Eulerian observer, in the EF, has
4-velocity $n^\mu = (1, 0, 0, 0)$.

\subsection{Frame transformations}
\label{sect:tetrad}

The tetrad
transformation provides the link between the CF and the EF. The matrix
$M$ is the transformation matrix, so that, for the basis vectors,
\begin{equation}
  \label{eq:dte}
  \vec{\partial}_{\tilde{\mu}} = M^\nu\,\!_{\tilde{\mu}} \, \vec{e}_\nu~,
\end{equation}
and
\begin{equation}
  \label{eq:etd}
  \vec{e}_\nu = M^{-1}\,\!\!_\nu\,\!^{\tilde{\mu}} \, \vec{\partial}_{\tilde{\mu}}~.
\end{equation}

An alternative definition is to say that for a given metric
$g_{\tilde{\mu}\tilde{\nu}}$, there always exists a matrix $M$ such
that
\begin{equation}
  g_{\tilde{\mu}\tilde{\nu}} = M^{\rho}\,\!_{\tilde{\mu}}
  M^{\gamma}\,\!_{\tilde{\nu}} \eta_{\rho \gamma}~,
\end{equation}
following, for example \cite{card_05, ende_12c}. So, the transformation
from the CF to the EF is performed by applying the transformation
matrix $M$ or its inverse (in the usual sense of a matrix inverse)
$M^{-1}$.

The line element in $3+1$ is given by
\begin{equation}
  \label{eq:lineelem}
  \d{}s^2 = -\alpha^2 \d{}t^2 + \gamma_{\tilde{i}\tilde{j}}(\d{}x^{\tilde{i}} + \beta^{\tilde{i}} \d{}t)(\d{}x^{\tilde{j}} + \beta^{\tilde{j}} \d{}t)~,
\end{equation}
where $\alpha$ is the lapse function, $\beta^{\tilde{i}}$ the shift
vector and $\gamma_{\tilde{i}\tilde{j}}$ the spatial 3-metric. With
this definition of $\alpha$ and $\beta^{\tilde{i}}$, we can
explicitly compute $M^{-1}\,\!\!_0\,\!^{\tilde{\nu}}$,
$M^{-1}\,\!\!_\mu\,\!^{\tilde{0}}$ and
$M^{-1}\,\!\!_0\,\!^{\tilde{0}}$. We have $\vec{e}_0 = \vec{n}$,
since $\vec{n}$ is unitary and future pointed. From the $3+1$
decomposition of $\vec{\partial}_0$ \cite{3plus1}
\begin{equation}
  \vec{e}_0 = \frac{1}{\alpha} \left( \vec{\partial}_0 - \beta^{\tilde{i}}
    \vec{\partial}_{\tilde{i}} \right)~,
\end{equation}
by identification with the tetrad transformation $\vec{e}_0 =
M^{-1}\,\!\!_0\,\!^{\tilde{\mu}} \, \vec{\partial}_{\tilde{\mu}}$, we
see that
\begin{equation}
  M^{-1}\,\!\!_0\,\!^{\tilde{0}} = \frac{1}{\alpha} \quad \textrm{and} \quad
  M^{-1}\,\!\!_0\,\!^{\tilde{i}} = - \frac{\beta^{\tilde{i}}}{\alpha}~.
\end{equation}

Then, we can express the spatial part of Eq.~(\ref{eq:dte}) for the CF basis
vectors as,
\begin{equation} 
 \vec{\partial}_{\tilde{i}} = M^\nu\,\!_{\tilde{i}} \, \vec{e}_\nu~.
\end{equation}
It becomes obvious that the $\vec{\partial}_{\tilde{i}}$ are functions of both
$\vec{e}_i$ and $\vec{e}_0$. However, both $\vec{\partial}_{\tilde{i}}$ and
$\vec{e}_i$ are spacelike and are tangent to the hypersurfaces of constant
time coordinate. If we also consider that $\vec{e}_0$ is timelike, we know
that $\vec{\partial}_{\tilde{i}}$ cannot depend on
$\vec{e}_0$. This implies that
\begin{equation}
  \label{eq:vierbzero}
  M^0\,\!_{\tilde{i}} = 0~.
\end{equation}
Since we know that $M M^{-1} = I$, we can infer
the general form of $M^{-1}$ in $3+1$,
\begin{equation}
  \label{eq:vierb3p1}
  M^{-1}\,\!\!_\mu\,\!^{\tilde{\nu}} = \begin{pmatrix}[2]
\dfrac{1}{\alpha}, & 0, &
0, & 0, \\ 
-\dfrac{\beta^r}{\alpha},  \\
-\dfrac{\beta^\theta}{\alpha}, & {} & M^{-1}\,_i\,^{\tilde{j}} \\
-\dfrac{\beta^\phi}{\alpha},
\end{pmatrix},
\end{equation}
where $\mu$ is the column index and $\tilde{\nu}$ is the row index
(see also \cite{ende_12c}). $M^{-1}\,_i\,^{\tilde{j}}$ is not uniquely
defined (it is possible to incorporate a Lorentz boost and/or a
rotation in it). We formulate it in section~\ref{sec:cfc} for our
choice of gauge, which is closely related.

We choose to express the general relativistic Liouville operator,
Eq.~(\ref{eq:b1}), in the EF. So, the parts of this equation that
originally belong to the CF, namely the spacetime derivatives and the
Christoffel symbols, have to be transformed to the EF.

The Eulerian time derivative is
\begin{equation}
  \label{eq:eulertime}
  D_t = M^{-1}\,\!\!_0\,\!^{\tilde{\mu}} \, \partial_{\tilde{\mu}} =
  \frac{1}{\alpha}\frac{\partial}{\partial t} -
  \frac{\beta^{\tilde{i}}}{\alpha}\frac{\partial}{\partial x^{\tilde{i}}}~,
\end{equation}
with the use of Eq.~(\ref{eq:vierb3p1}).

The spatial derivatives in the Eulerian frame are

\begin{equation}
  \label{eq:eulerspatial}
  D_i = M^{-1}\,\!\!_i\,\!^{\tilde{\mu}} \, \partial_{\tilde{\mu}} =
  M^{-1}\,\!\!_i\,\!^{\tilde{j}} \, \partial_{\tilde{j}}~,
\end{equation}
where the last step comes from the fact that $M^{-1}\,_i\,^0 = 0$, so
that we replace $\tilde{\mu}$ with a spatial index. 

The momentum 4-vector in the supernova neutrino transport problem is
easily defined by its components in the LF, because we use the 4-momentum
to calculate the collision operator. In principle, in
Eq.~(\ref{eq:b1}) each momentum and derivative with respect to the
momentum component has to be transformed to the EF. With the LF and EF
definition, the only difference between these two frames is the
3-velocity of the fluid. So, the transformation between both is a
Lorentz boost
\begin{equation}
  \label{eq:boost}
  p^\nu = \Lambda^\nu\,\!_{\hat{\mu}} p^{\hat{\mu}}.
\end{equation}
Note that the Lorentz boost depends on the Eulerian 3-velocity of the
fluid $v^i$. 
We choose, for this article, to consider only static
configurations. The main reason for this choice is because our
radiative transfer code is not yet coupled to a hydrodynamics
code. This choice implies that the velocity $v^i$ vanishes and that
the momentum in the EF is the same as in the LF.

\subsection{Ricci rotation coefficients}
\label{sec:riccicoeffs}

\subsubsection{Definition\\}

Ricci rotation coefficients are connection coefficients. One can
define them with the covariant derivative $\nabla$ associated with the
metric $\bm{g}$.

In the CF, the covariant derivative along a basis vector
$\vec{\partial}_{\tilde{\nu}}$ of another basis vector
$\vec{\partial}_{\tilde{\mu}}$ is 
\begin{equation}
  \nabla_{\tilde{\nu}} \vec{\partial}_{\tilde{\mu}} = \Gamma^{\tilde{\rho}}\,\!_{\tilde{\mu} \tilde{\nu}} \, \vec{\partial}_{\tilde{\rho}}~,
\end{equation}
where the connection coefficients on the CF
$\Gamma^{\tilde{\rho}}\,\!_{\tilde{\mu} \tilde{\nu}}$ are the Christoffel symbols.

Analogously, on the EF, the covariant derivative along a basis vector
$\vec{e}_{\nu}$ of another basis vector $\vec{e}_{\mu}$ is
\begin{equation}
  \label{eq:defricci}
  \nabla_{\nu} \vec{e}_\mu = \bar{\Gamma}^\rho\,\!_{\mu \nu} \, \vec{e}_{\rho}~,
\end{equation}
where the connection coefficients on the EF
$\bar{\Gamma}^\rho\,\!_{\mu \nu}$ are different from the Christoffel
symbols. They are what we call Ricci rotation coefficients.

Connection coefficients are not tensors, and need a special treatment
which we describe here (see also,
\textit{e.g.}~\cite{card_05}). First, we introduce a general
formulation, then the $3+1$ formalism, and finally the conformally
flat approximation.

\subsubsection{General formulation\\}

Because the Christoffel symbols do not define a tensor, their transformation
to the EF is more complicated. One can link the Ricci rotation coefficients
with the Christoffel symbols (see {\em{e.g.}}~\cite{card_13}), but we choose to
express them in a way that is more convenient for us. However, unlike the
Christoffel symbols that are symmetric in the last two indices, the Ricci
rotation coefficients are antisymmetric with respect to the two first indices
\begin{equation}
  \label{eq:antisym}
  \bar{\Gamma}^{\mu}\,\!_{\nu \rho} = - \bar{\Gamma}^{\nu}\,\!_{\mu \rho}.
\end{equation}

One can define the tetrad derivative as
\begin{equation}
  \label{eq:vierbderiv}
  d^\rho\,\!_{\mu \nu} = - M^{-1}\,\!\!_\mu\,\!^{\tilde{\lambda}} \,
  M^{-1}\,\!\!_\nu\,\!^{\tilde{\kappa}} \, \frac{\partial
    M^\rho\,\!_{\tilde{\lambda}}}{\partial x^{\tilde{\kappa}}}.
\end{equation}
Like the Christoffel symbols, the tetrad derivative, $d^\rho\,_{\mu
  \nu}$, is not a tensor. The ``no torsion'' condition of GR is expressed as
\begin{equation}
  \label{eq:notors}
  \bar{\Gamma}^{\mu}\,\!_{\nu \rho} - d^{\mu}\,\!_{\nu \rho} = \bar{\Gamma}^{\mu}\,\!_{
    \rho \nu} - d^{\mu}\,\!_{\rho \nu}.
\end{equation}
With Eqs.~(\ref{eq:antisym}) and (\ref{eq:notors}), and considering
the definitions
\begin{eqnarray}
\bar{\Gamma}_{\mu \nu \rho} = \eta_{\mu \alpha}\bar{\Gamma}^{\alpha}\,\!_{\nu \rho},\\
d_{\mu \nu \rho} = \eta_{\mu\alpha} d^{\alpha}\,\!_{\nu \rho},
\end{eqnarray}
we find a convenient way to rewrite the Ricci rotation coefficients

\begin{equation}
  \label{eq:computericci}
  \bar{\Gamma}_{\mu \nu \rho} = \frac{1}{2} \left[ d_{\mu \nu \rho} -
    d_{\mu \rho \nu} + d_{\nu \rho \mu} - d_{\nu \mu \rho} +
    d_{\rho \nu \mu} - d_{\rho \mu \nu} \right].
\end{equation}

\subsubsection{$3+1$ formulation eliminating the time derivative\\}
The $3+1$ Ricci rotation coefficients we need to express are
$\bar{\Gamma}^{i}\,\!_{j k}$, $\bar{\Gamma}^{i}\,\!_{0 0}$,
$\bar{\Gamma}^{i}\,\!_{0 j}$ and $\bar{\Gamma}^{i}\,\!_{j 0}$. Expressing
them with the first index raised or lowered is completely equivalent.

$\bar{\Gamma}_{i j k}$ is computed with Eq.~(\ref{eq:computericci}). A
consequence of Eq.~(\ref{eq:vierbzero}) is that fully spatial Ricci
rotation coefficients imply that $d$ are fully spatial too,
\textit{i.e.}, there is no time derivative in the expression
(Eq.~(\ref{eq:vierbderiv})) of $d_{i j k}$.

$\bar{\Gamma}_{i 0 0}$ is computed with
Eq.~(\ref{eq:computericci}) and reduces to
\begin{equation}
  \bar{\Gamma}_{i 0 0} = d_{0 0 i},
\end{equation}
where
\begin{equation}
  d_{0 0 i} = - \eta_{0 0} M^{-1}\,\!\!_0\,\!^{\tilde{\mu}} \,
  M^{-1}\,\!\!_i\,\!^{\tilde{\nu}} \, \frac{\partial
    M^0\,\!_{\tilde{\mu}}}{\partial x^{\tilde{\nu}}} = \frac{1}{\alpha}
  M^{-1}\,\!\!_i\,\!^{\tilde{j}} \, \frac{\partial \alpha}{\partial x^{\tilde{j}}}~.
\end{equation}

To compute $\bar{\Gamma}_{i 0 j}$, we use Eq.~(\ref{eq:defricci})
\begin{equation}
  \bar{\Gamma}_{i 0 j} = \vec{e}_i \nabla_j \vec{e}_0~.
\end{equation}
This is by definition the opposite of the extrinsic curvature $K_{ij}$
(see \cite{3plus1}), the sign being conventional (this convention
agrees with \cite{3plus1}, \cite{cord_09}).

\begin{equation}
  \bar{\Gamma}_{i 0 j} = -K_{i j}.
\end{equation}

Finally, $\bar{\Gamma}_{i j 0}$ is computed by applying
Eq.~(\ref{eq:notors}), so that

\begin{equation}
  \label{eq:gammaij0}
  \bar{\Gamma}_{i j 0} = \bar{\Gamma}_{i 0 j} - d_{i 0 j} + d_{i j 0}.
\end{equation}
The calculation of $d_{i 0 j}$ and $d_{i j 0}$ requires some algebra
that is shown in detail in \ref{app:di0jdij0}.

\subsubsection{Conformally flat formulation\\}
\label{sec:cfc}

In the conformally flat approximation (also called conformally flat
condition or CFC, see~\cite{wils_96}), $\gamma_{\tilde{i}\tilde{j}}$
in the line element Eq.~(\ref{eq:lineelem}) is
\begin{equation}
  \label{eq:defcfc}
  \gamma_{\tilde{i}\tilde{j}} = \Psi^4 f_{\tilde{i}\tilde{j}},
\end{equation}
where $f_{\tilde{i}\tilde{j}}$ is the flat metric and $\Psi$ is the conformal
factor:
\begin{equation}
  \label{e:def_Psi}
  \Psi = \left( \frac{\det \gamma_{\tilde{i}\tilde{j}}}{\det
      f_{\tilde{i}\tilde{j}}} \right)^{1/12}.
\end{equation}

In the CFC, a straightforward choice for the spatial part of the tetrad
transformation matrix $M$ is a diagonal matrix (or equivalently, the
$\vec{\partial}_{\tilde{i}}$ basis vectors are aligned with the
$\vec{e}_i$). Finally, in spherical coordinates we get
\begin{equation}
  \label{eq:vierbcfc}
  M^{-1}\!\!\,_\mu\,\!^{\tilde{\nu}} = \begin{pmatrix}[2]
      \dfrac{1}{\alpha}, & 0, & 0, & 0, \\ 
      -\dfrac{\beta^r}{\alpha}, & \dfrac{1}{\Psi^2}, & 0, & 0, \\
      -\dfrac{\beta^\theta}{\alpha}, & 0, & \dfrac{1}{\Psi^2 r}, & 0, \\
      -\dfrac{\beta^\phi}{\alpha}, & 0, & 0, & \dfrac{1}{\Psi^2 r \sin \theta},\\
\end{pmatrix}
\end{equation}
and
\begin{equation}
  M^\mu\,_{\tilde{\nu}} = \begin{pmatrix}[2]
      \alpha, & \Psi^2 \beta^r, & \Psi^2 r \beta^\theta, & \Psi^2 r
      \sin \theta \beta^\phi, \\
      0, & \Psi^2, & 0, & 0, \\
      0, & 0, & \Psi^2 r, & 0, \\
      0, & 0, & 0, & \Psi^2 r \sin \theta
\end{pmatrix}.
\end{equation}
Since the fully spatial part of $M$ and $M^{-1}$ are diagonal, this
approximation naturally simplifies the calculation of the Ricci
rotation coefficients. The CFC is a suitable choice for modelling
supernova phenomena (\textit{e.g.}~\cite{shib_04, cerd_05}) although
it misses some aspects of GR. The CFC does not contain gravitational
waves, and it cannot exactly describe a Kerr black hole or a rotating
fluid configuration, which might affect some simulations of rotating black hole
formation, as in hypernovae. Even in that case, it is not obvious that
the CFC would be a major source of inconsistency, as it captures many
non-linear features of GR (\textit{e.g.} non-rotating black holes are
exactly described) and can even include some rotational properties of
the spacetime.

\subsection{Boltzmann equation in the CFC}

With all the tools introduced in the previous sections, we get to our
formulation of the Boltzmann equation, with the Liouville operator in the EF,
no velocity dependent terms and within the CFC 
\begin{equation}
  \label{eq:boltzefcfc}
  \frac{1}{\alpha}\frac{\partial f}{\partial t} + \left(
    \frac{p^i}{\Psi^2 \epsilon} -
  \frac{\beta^i}{\alpha} \right) \frac{\partial f}{\partial x^i} -
\bar{\Gamma}^j\,\!_{\mu \nu} p^\mu p^\nu J^i\,\!_j  \frac{1}{\epsilon}\frac{\partial f}{\partial p^i} = \frac{1}{\epsilon}\mathcal{C}[f],
\end{equation}
which is the equation used in the numerical tests. Note that the
Eulerian spatial derivatives, Eq.(\ref{eq:eulerspatial}), in the CFC, only
implies to multiply by an additional prefactor $1 / \Psi^2$ with
respect to the CF spatial derivatives
(Eq.(\ref{eq:realspaceop})). Note also that no distinction is made
between LF, CF and EF quantities in Eq.~(\ref{eq:boltzefcfc}), as they
would obscure the point.

Let us insist on the fact that the EF formulation presented here is only a
matter of convenience. When looking at Eq.~(\ref{eq:boltzeffull}), one can see
that $f$ is a function of $t$, $x^{\tilde{i}}$ (space-time coordinates in the
CF) and $p^{\hat{k}}$ (momentum space coordinates in the LF). This is the
usual dependence of $f$ in the study of neutrinos in supernovae
(\textit{e.g.}~\cite{mezz_89, muel_10, card_05}). Moreover, $L[f]$ and
$\mathcal{C}[f]$ being Lorentz invariants, it is still possible to compute
$\mathcal{C}[f]$ in the LF, which is by far the most convenient for this
operator.

\section{Test cases}\label{sec:testresults}

The \texttt{Ghost} numerical code  is detailed in
\ref{app:implementation}; in this section we discuss the different
tests we applied to the {\tt{Ghost}} code to ensure the validity and
reliability of our numerical results. We treated the nonrelativistic
methods separately from the relativistic methods, and thus discuss
them separately below.

\subsection{Convergence tests}

To test the convergence of the 6-dimensional spectral representation
in all dimensions we use exponentials, which represent infinite series
of polynomials or Fourier series, enabling a test of truncation
error. For a given dimension we want to test, represented by a
coordinate $x$, the test proceeds as follows

\begin{itemize}
\item We represent a function $g = \exp(0.1 x)$ for $x = r, \epsilon,
  \Theta$ or $g = \exp(\cos(x))$ for $x = \theta, \phi, \Phi$.
\item We apply the operator $\partial /\partial x$.
\item We compare the numerical value at a given point $(r_0, \theta_0, \phi_0, \epsilon_0, \Theta_0,
  \Phi_0)$ to the analytic value of the derivative at the same point.
\end{itemize}

The resolution used is $n_r = 9$, $n_\theta = 9$, $n_\phi = 4$,
$n_\epsilon = 9$, $n_\Theta = 9$, $n_\Phi = 4$ and we vary only the
number of points in the tested dimension.
One radial shell is simulated, from $R_{\min} = 1 \ \textrm{km}$ to
$R_{\max} = 4 \ \textrm{km}$, as well as one energy shell, from
$\epsilon = 10 \ \textrm{MeV}$ to $\epsilon = 30 \
\textrm{MeV}$. The choices for the radial shell and energy are entirely arbitrary. 
The $\Theta$ interval is divided in two domains, $0 \le \Theta \le \pi/2$ 
and $\pi/2 \le \Theta \le \pi$, as discussed in \ref{app:time}. 
Both domains have $n_\Theta$ points.

The arbitrary point where we evaluate both functions is $(r_0, \theta_0, \phi_0,
\epsilon_0, \Theta_0, \Phi_0) = (1.2, 0.3, 0.3, 10.3, 1.45, 0.3)$; the
important fact being that it is not a grid point (\textit{i.e.} not a
collocation point). 

We calculate the relative error by,
\begin{equation}
  \label{rel_err_6d}
 {\rm{R.E.}} = \frac{\left| \left.\frac{\partial g}{\partial
         x}\right|_{c} - \left.\frac{\partial g}{\partial
         x}\right|_{a} \right|}{\left.\frac{\partial g}{\partial
       x}\right|_{a}},
\end{equation}
where $\partial g/\partial x |_{c}$ denotes the numerically calculated
derivative evaluated at a specific coordinate, and $\partial
g/\partial x |_{a}$ denotes the closed form of the derivative
evaluated at the same coordinate. As expected, we observe an exponential
decay of this relative error for each of the six dimensions tested,
down to the numerical round-off error ($\sim 10^{-14}$). This floor is
reached with less than 20 coefficients in the considered argument,
with the exception of $\phi$ and $\Phi$ arguments, where we need
about 30 coefficients.

\begin{figure}
  \leftskip -2.5cm
  {
  \begin{tabular}{cc}
    \noindent
    \includegraphics[width = .65\textwidth]{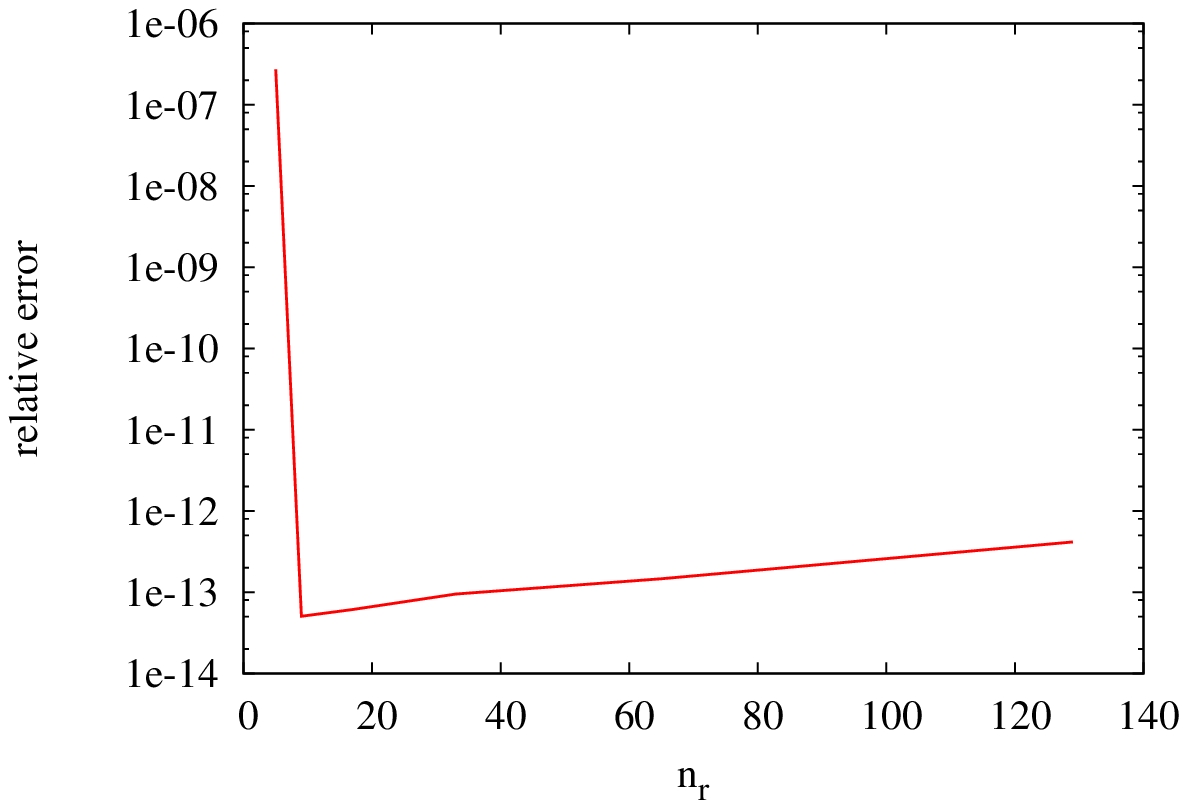}      
    \includegraphics[width = .65\textwidth]{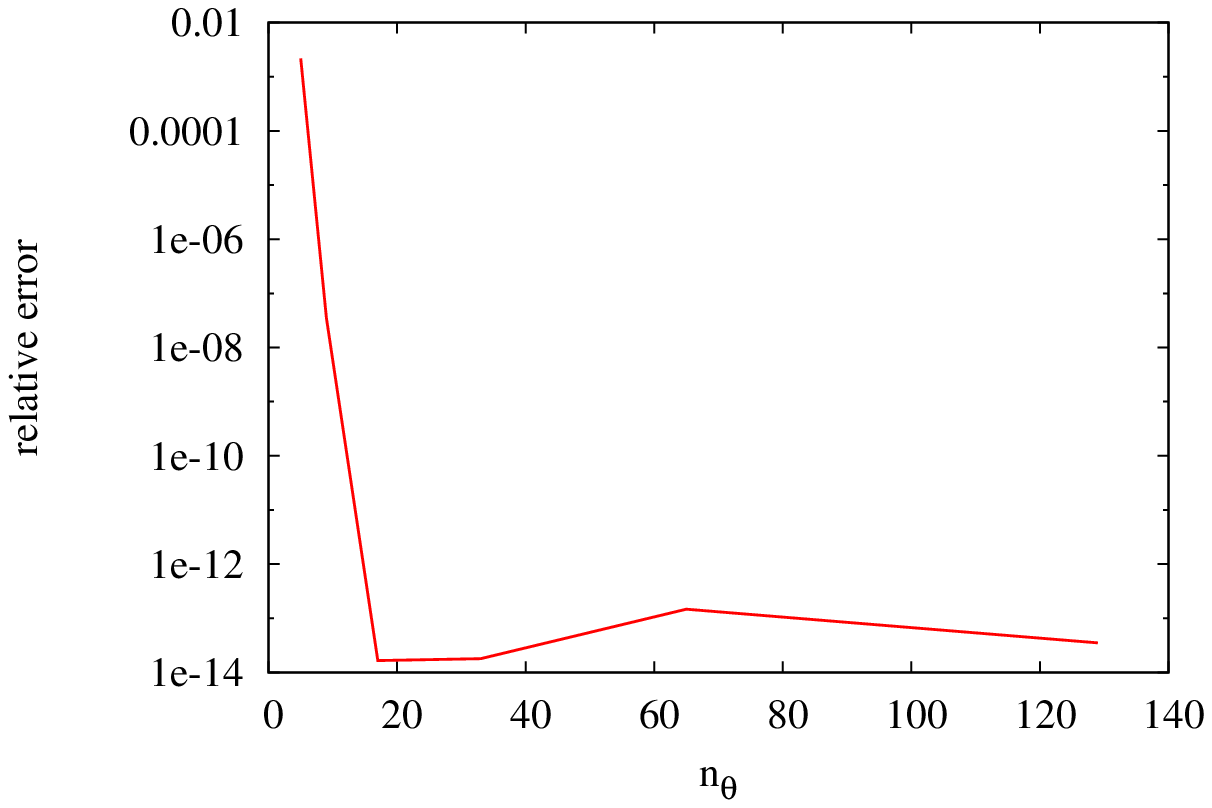}    \\
    \includegraphics[width = .65\textwidth]{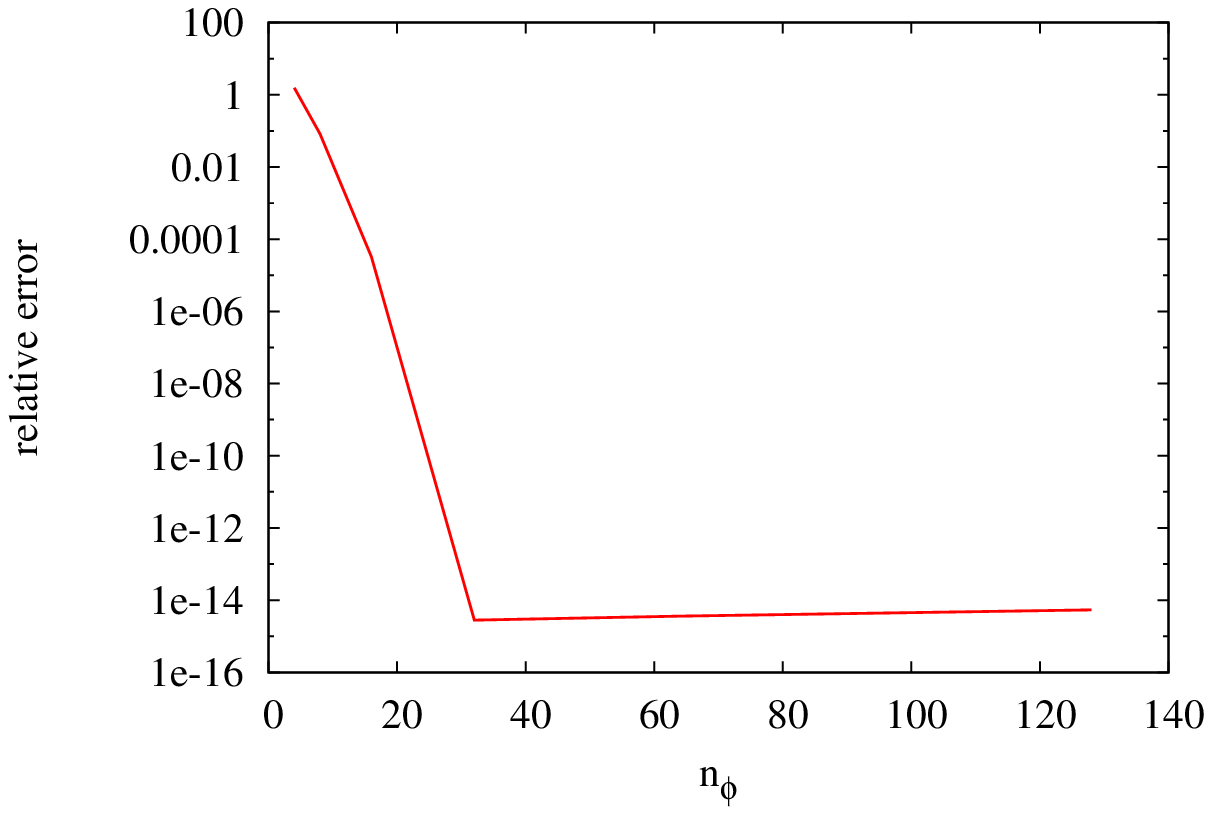}      
    \includegraphics[width = .65\textwidth]{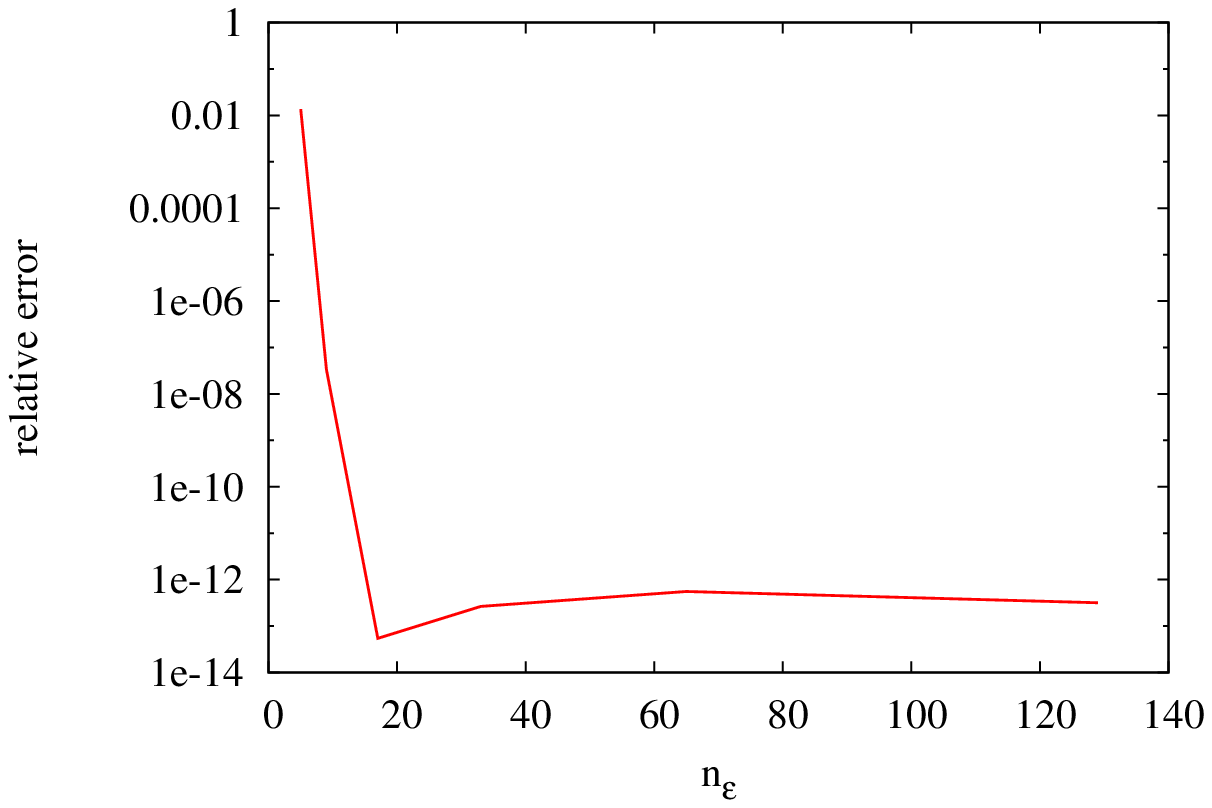}    \\
    \includegraphics[width = .65\textwidth]{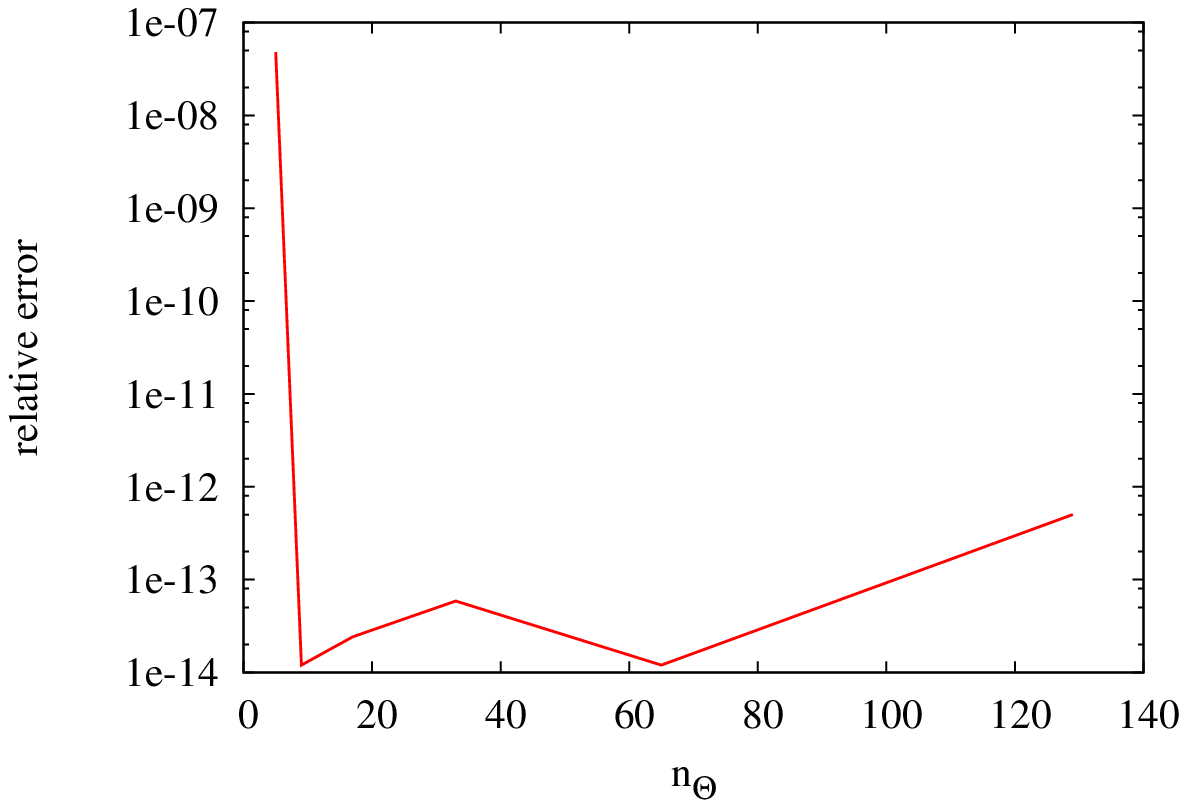}      
    \includegraphics[width = .65\textwidth]{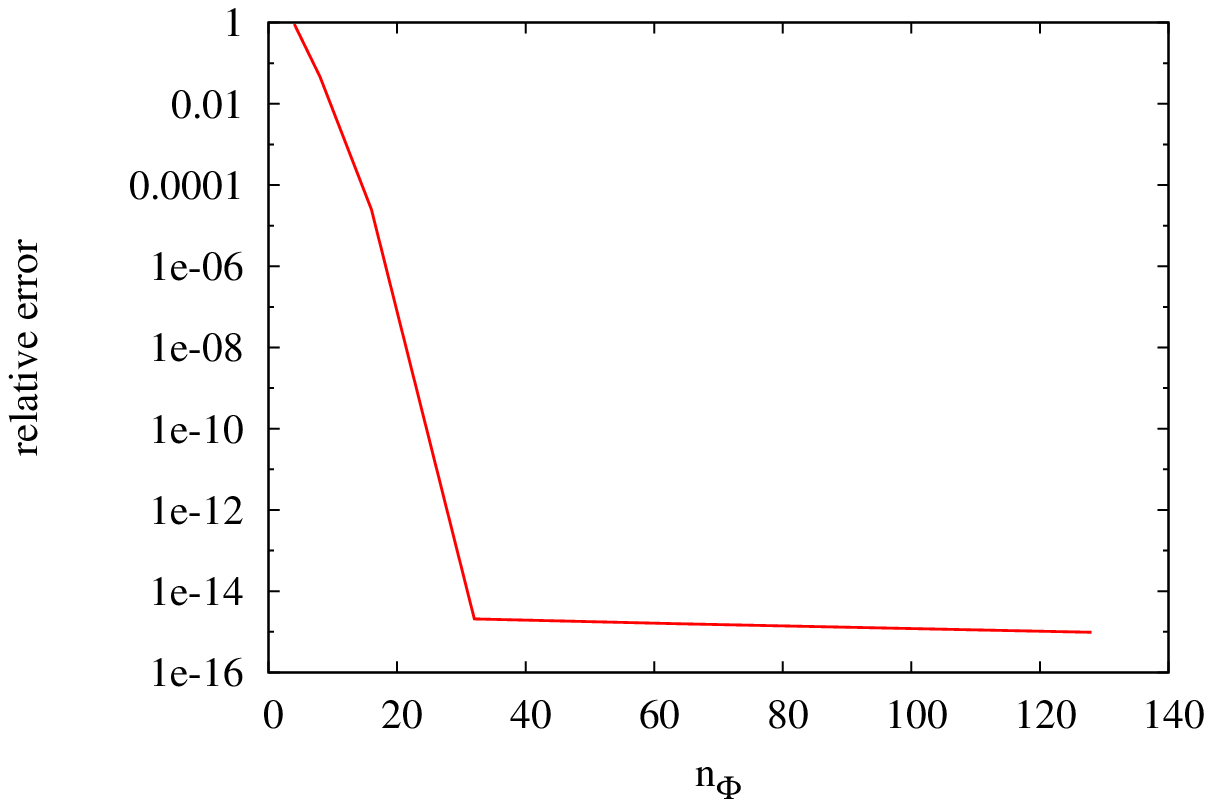}    \\
  \end{tabular}
}
\caption{Relative error R.E., Eq.~(\ref{rel_err_6d}) as a function of
  the number of points in the tested dimension, for each of them ($r$,
  $\theta$, $\phi$, $\epsilon$, $\Theta$, $\Phi$). We can see a
  spectral (i.e., exponential) convergence in every case.}
  \label{fig:converg_all}
\end{figure}

\subsection{Collision term}
The full collision operator is not yet implemented in our code. For
the test cases we implement an isoenergetic scattering
\begin{equation}
  \label{e:collision}
  B_{IS} = \frac{\epsilon^2}{(2 \pi)^3} \int_0^\pi
  \int_0^{2 \pi} R_{IS} \sin (\Theta') \left[ f(r, \theta, \phi,
    \epsilon, \Theta', \Phi') - f(r, \theta, \phi, \epsilon, \Theta,
    \Phi) \right] \d{}\Theta' \d{} \Phi',
\end{equation}
with $R_{IS}$ the scattering kernel (see \cite{brue_85}). We choose to
implement the scattering over nucleons (neutrons and protons), for which we
follow \cite{brue_85}. This term is the only contribution to the
collision operator implemented in the test cases.

An interesting point is that the collision term is often treated using
a decomposition on Legendre polynomials, keeping only the first two
terms (\textit{e.g.}~\cite{brue_85, ramp_02}). Thus, it is possible to
compare the spectral treatment using Legendre polynomials to our
methods, knowing that in our case it is very easy to improve
accuracy. This shall be done in a forthcoming study. There may not be
very large differences between our method and the Legendre polynomial
method, but in our opinion this is interesting to check.

\subsection{Particle number conservation in GR}
Here, for a given initial condition, we evolve the Boltzmann equation
$L[f] = \mathcal{C}[f]$ in time and look whether the total number of
neutrinos is conserved by our numerical scheme. The (time-dependent)
number of particles is computed via
\begin{equation}
  N(t) = \frac{1}{(2 \pi)^3} \int f (t, r, \theta, \phi, \epsilon, \Theta,
  \Phi) \d^6V~,
\end{equation}
and the relative particle number conservation is computed as
\begin{equation}
  \label{eq:relativenbpartic}
  \delta = \frac{N^{n} - N^{0}}{N^{0}}~,
\end{equation}
where $N^{n}$ is the number of particles at the $n$-th time step, and $N^{0}$
is the number of particles at the beginning of the simulation. $\delta$ is
also corrected for the radial fluxes at the boundaries.

One radial shell is simulated, from $R_{\min} = 1 \ \textrm{km}$ to $R_{\max}
= 2.5 \ \textrm{km}$. The boundary conditions are as follows: $f (R_{\min},
\theta, \phi, \epsilon, \Theta < \pi / 2, \Phi)$ is held constant (equal to
the initial condition), and $f (R_{\max}, \theta, \phi, \epsilon, \Theta > \pi
/ 2, \Phi)=0$ and is also held constant.  The time step is constant
(see~\ref{app:time} for an explanation of its calculation), and set to
\begin{equation}
 \Delta t = 0.1 \frac{1}{n_\Theta^2}.
\end{equation}
The initial conditions used are as follows,
\begin{equation}
f (r, \theta, \phi, \epsilon, \Theta, \Phi, t=0) = \left(
  \frac{p_Z}{\epsilon} \right)^2 \times H \left(
    \Theta-\frac{\pi}{2} \right)~,
\end{equation}
where $H(\Theta-\pi/2)$ is the Heaviside function. $p_Z$ is the
projection of the momentum vector on the corresponding Cartesian axis
(see~\ref{app:transfomatrix} for the detailed expression of this quantity).

In this test we evolve the Boltzmann equation
Eq.~(\ref{eq:boltzefcfc}) with the general relativistic Liouville
operator, and use the collision operator that includes isoenergetic
scattering where the value of the nucleon density $n_b = 2 \times
10^{13} \ \textrm{g.cm}^{-3}$ corresponds to an opaque regime . We do
not make any assumptions about the symmetry of the system. We use $9
\times 9 \times 4 \times 5 \times 33 \times 4$ points in $r \times
\theta \times \phi \times \epsilon \times \Theta \times \Phi$.

To simplify matters we evolve our Boltzmann equation on a
Schwarzschild spacetime background\footnote{It should be noted that
  more complicated spacetime backgrounds are possible, but would
  obscure the point of the tests. Moreover, due to the simplicity of
  this metric, the CFC is not an approximation in this case.}.
The Schwarzschild metric is written in isotropic gauge 
\begin{equation}
  \label{e:def_Schwarz}
  \d{}s^2 = - \left( \frac{1 - \frac{R_s}{4 r}}{1 + \frac{R_s}{4 r}}
  \right)^2 \d{}t^2 + \left( 1 + \frac{R_s}{4 r} \right)^4 (\d{}r^2 +
  r^2 \d{} \theta^2 + r^2 \sin^2 \theta \d{} \phi^2),
\end{equation}
with $R_s = 2M$ the Schwarzschild radius. This expression is nothing but
the CFC expression of the line element in Eqs.~(\ref{eq:lineelem}) and
(\ref{eq:defcfc}), with
\begin{equation}
  \alpha = \dfrac{1 - \dfrac{R_s}{4 r}}{1 + \dfrac{R_s}{4 r}},
  \quad \quad \beta^i = 0 \quad \textrm{and} \quad \Psi = 1
  + \frac{R_s}{4 r}.
\end{equation}
In our simulations, we use $R_s = 0.4 \ \textrm{km}$ so that the
horizon is not within the computational domain.

\begin{figure}
  \centering
  \input{new_conserv_relat_15.tex}
  \caption{Particle conservation relative error as a function of time.}
  \label{fig:partic_cons}
\end{figure}
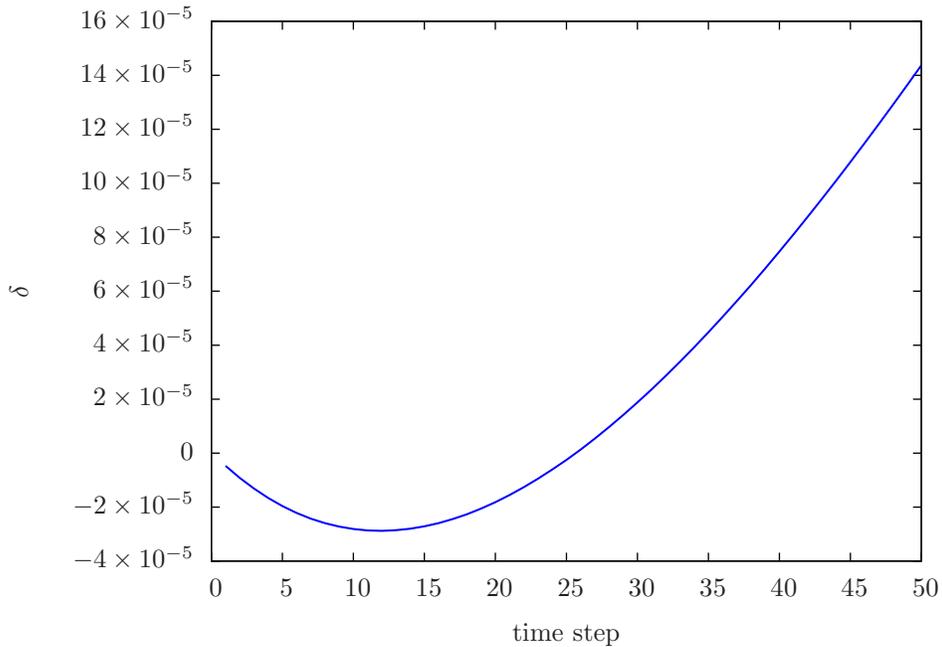

The particle number conservation is at most $\delta = 10^{-4}$ all along the
50 simulated time steps (see Fig.~\ref{fig:partic_cons}). We checked that it
is converging with decreasing the time step. This provides evidence that the
time evolution is working and our calculation conserves the number of
particles, with a small numerical error.

\subsection{Gravitational redshift test}

For this test, we again use the Schwarzschild metric and attempt to
recover the expected gravitational redshift.

The physical setup of this test is described as follows. 
Consider a static observer at $r = R_{\rm min}$ that compares a neutrino
spectrum he sees with the one of a static observer at $r = R_{\rm max}$.

The static observer is, by definition, the Eulerian observer. However, to
compare the two spectra, we need to transform the observations into the CF,
because it is the only global frame (recall that the EF at $R_{\rm min}$ is
different from the EF at $R_{\rm max}$). In the EF, the 4-velocity is always
$n^\mu = (1, 0, 0, 0)$ and so a neutrino spectrum, as seen in the EF, would
not experience any redshift.

In the CF, we have the usual formulation
\begin{equation}
  \label{eq:redshift}
  \frac{\epsilon^{CF}_{2}}{\epsilon^{CF}_{1}} = \frac{\alpha_1}{\alpha_2},
\end{equation}
where $1$ denotes the emission point and $2$ denotes the reception point. This
equation can also be found by applying $M^{-1}$ to $n^\mu$.

\begin{figure}
  \centering
  \input{redshift_test_cf_smooth.tex}
  \caption{Distribution function energy profiles in Schwarzschild
    spacetime~(\ref{e:def_Schwarz}), as seen by static observers. The
    profile at $r=R_{\rm min}$ (red, plain) is imposed, the profile at
    $r=R_{\rm max}$ (green, dashed) is then deduced after integration
    of the neutrino transport equation in vacuum. The observer
    4-velocity has been transformed back to the CF for clarity.}
  \label{fig:redshift}
\end{figure}
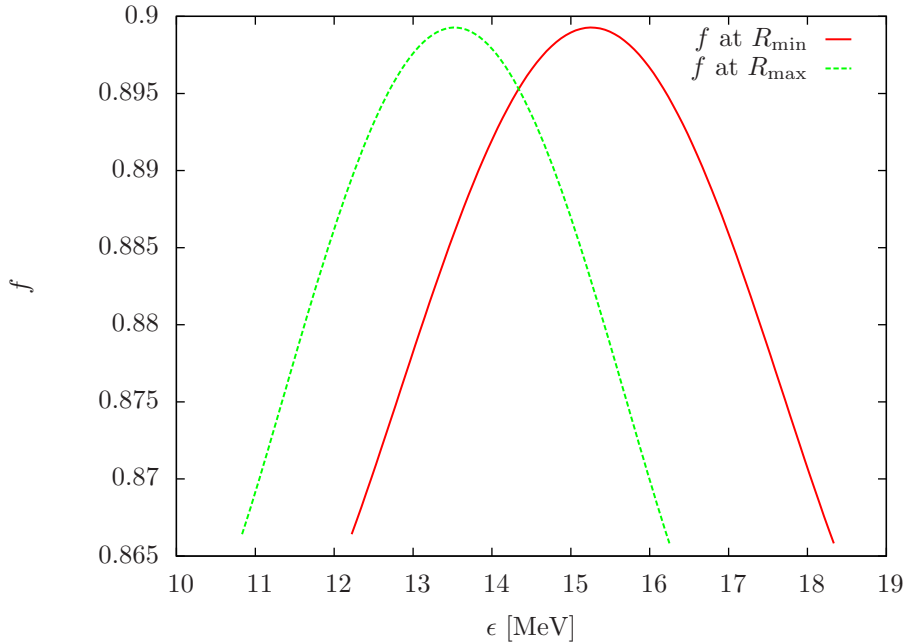

The numerical test consists on simulating a radial shell with $R_{\rm
  min} = 5M$ and $R_{\rm max} = 12.5M$. We use a resolution $n_r = 9$,
$n_\theta = 5$, $n_\phi = 4$, $n_\epsilon = 9$, $n_\Theta = 5$,
$n_\Phi = 4$. With initial condition $f = 0$ everywhere, except at the
inner boundary ($r = R_{\rm min}$) where a flux of neutrinos is
injected with a Gaussian profile in $\epsilon$. By solving the
Boltzmann equation (without collisions, and without any assumption on
symmetries), we recover the same Gaussian profile at $r = R_{\rm max}$
(see figure~\ref{fig:redshift}). We then transform the spectra to the
CF to be able to compare them and to show the gravitational redshift.

As we see in figure~\ref{fig:redshift}, in the CF, the spectrum is
redshifted when travelling from $R_{\rm min}$ to $R_{\rm max}$, by a factor
$\alpha_1 / \alpha_2 = 0.89$.

With this configuration (GR with no collision), a time step takes
approximately one second. When improving the resolution to $n_r = 17$,
$n_\theta = 9$, $n_\phi = 4$, $n_\epsilon = 17$, $n_\Theta = 17$, $n_\Phi =
4$, a time step takes approximately 23 seconds. Improving the resolution again
by a factor of 2 with $n_r = 17$, $n_\theta = 9$, $n_\phi = 4$, $n_\epsilon =
17$, $n_\Theta = 33$, $n_\Phi = 4$, a time step take approximately 47
seconds. (tests done with a single processor Intel(R) Core(TM) i7-3630QM CPU
with a frequency of 2.4 GHz).


\subsection{Searchlight beam test}

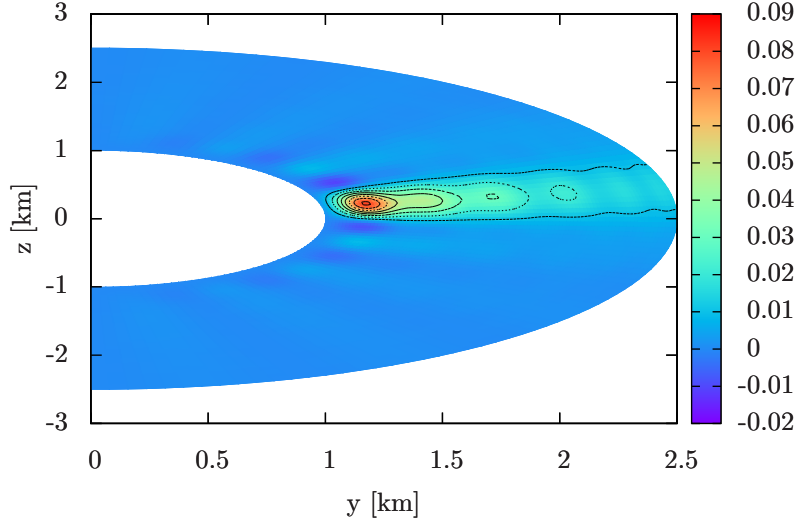
\begin{figure}
  \centering
  \input{searchlight.tex}
  \caption{Searchlight beam test. The colors and the black contour lines code
    the value of the neutrino intensity, in an arbitrary $(r, \theta)$
    plane. The spread of the beam is mostly due to the initial spread of the
    source, rather than numerical errors.}
  \label{fig:searchlight}
\end{figure}

The searchlight beam test proposes to simulate the narrowest beam possible and
look at the dispersion due to numerical errors \cite{sumi_12, ston_92}. The
dispersion must be reduced by increasing resolution.

With spectral methods, introducing a point source would result in considerable
Gibbs phenomenon. When one tries to represent a non smooth (discontinuous)
function with spectral methods, unwanted oscillations appear. This is called
Gibbs phenomenon (see also \ref{app:implementation}). The bigger the
discontinuity, the bigger the oscillations, and if too big, the simulation may
give wrong and unphysical results.

Instead of introducing a point source, we introduce a Gaussian shaped source,
and try to get the narrowest Gaussian possible before have too much Gibbs
phenomenon. Indeed, at a given resolution, the representation of narrower and
narrower Gaussian leads to more and more Gibbs phenomenon, because a too
narrow Gaussian approaches a discontinuous function.

Initial conditions for this test are as follow. One radial shell is simulated,
from $R_{\min} = 1 \ \textrm{km}$ to $R_{\max} = 2.5 \ \textrm{km}$. We use a
resolution $n_r = 17$, $n_\theta = 17$, $n_\phi = 4$, $n_\epsilon = 1$,
$n_\Theta = 17$, $n_\Phi = 4$. We impose

\begin{equation}
  \label{eq:sourcef0}
  f_0 = \exp{-\frac{\left( r - \mu_r \right)^2}{2\sigma_r^2}} \times
  \exp{-\frac{\left( \theta - \mu_\theta \right)^2}{2\sigma_\theta^2}}  \times
  \exp{-\frac{\left( \Theta - \mu_\Theta \right)^2}{2\sigma_\Theta^2}}~,
\end{equation}

with $\mu_r = 1.2 \ \textrm{km}$, $\mu_\theta = 1.4$, $\mu_\Theta = 0.1$, and
$2\sigma_r^2 = 0.1 \ \textrm{km}^2$, $2\sigma_\theta^2 = 2\sigma_\Theta^2 =
0.1$.

Because the chosen value for $\Theta$ is very close to $0$, $\Phi$ only plays
a very minor role, so we do not multiply by a Gaussian in $\Phi$. Moreover,
nothing depends on energy in this test, so we completely ignore the energy
dependence ($n_\epsilon = 1$). Finally, for simplicity we evolve the $\phi$
degree of freedom but we do not multiply by a Gaussian in $\phi$, so that the
results can be plotted in 2D in a $r, \theta$ plane.

We evolve the Boltzmann equation without collision and without gravitation (as
it would obscure the point), with the source continuously emitting neutrinos
in the defined direction. We wait for the beam to cross the shell.

On figure \ref{fig:searchlight}, the result is shown in an arbitrary $(r,
\theta)$ plane. The momentum space dependence has been integrated. The color
codes the value of the distribution function (or, equivalently, the neutrino
intensity, because energy is constant), and we also plot black contour
lines. We see that the spreading is mostly due to the representation by
Gaussian profiles of the source, more than numerical errors. So, by
construction, increasing the resolution decreases the spread of the
source. Finally, on figure \ref{fig:searchlight}, we see that profiles that
narrow produce a small amount of unwanted Gibbs phenomenon. We note that this
test would benefit from a comparison with a finite differences code, and we
could give more quantitative results. We let this point for future work.

\section{Conclusion}
We have derived the general relativistic Boltzmann equation in doubly
spherical coordinates and with the Liouville operator in the Eulerian frame,
using a $3+1$ formalism. This derivation is particularly adapted for our
purposes, as it does not contain any time derivative, typically hard to tackle
with spectral methods. This formalism naturally avoids the singularity found
in the different parts of the Boltzmann equation. We have presented a $3+1$
version of the Liouville operator, and then the conformally flat
version. While we have worked in the CFC, the derivation presented here can
easily be generalized to 2D/3D cases in full general relativity, in other
gauge choices.

We have shown a first step towards numerically solving the general
relativistic neutrino transport problem in core-collapse supernovae
with spectral methods. The use of spectral methods, namely the
decomposition using a Chebyshev basis for the $r$, $\Theta$, and
$\epsilon$ coordinates, as well as the decomposition using a Fourier
basis for the $\theta$, $\phi$, and $\Phi$ coordinates, leads to a
substantial increase in numerical efficiency compared to existing
finite difference methods which are the current standard tool used for
the supernova neutrino problem. Ultimately, to reproduce the
resolution of the finite difference solutions, we require up to five
times fewer grid points per dimension. This fact allows us to solve
the neutrino transport problem using a single processor on a desktop
computer for the test cases, rather than super computers and a
substantial amount of memory. This work is a significant step towards
developing code that perform simulations of 3D supernova using 6D
Boltzmann neutrino transport in a reasonable amount of time

A possible improvement would be to implement a conservative version of
the Boltzmann equation (as discussed in~\cite{card_12, ende_12c}). In
these papers the authors demonstrate both the efficiency and the
improved accuracy of the conservative version with respect to the non
conservative version of the Boltzmann equation. The adaptation of the
conservative version to spectral methods represents a substantial
amount of work and remains to be done. The other direction of
improvement is the treatment of the central region in spherical
coordinates, which our code is not able to describe. This could be
done with a coupling of our code with a flux-limited diffusion one, as
we expect the center of the proto-neutron star to be opaque to the
neutrinos during post-bounce phase in core-collapse simulations.

It would be fruitful to perform a direct comparison between the
different methods used to approximate the Boltzmann equation. In this
proposed study we could test the accuracy of different approximations,
such as the momentum closure, the flux limited diffusion or the
ray-by-ray approximation. However, it is not a straightforward task to
link the use of moments in~\cite{shib_11} to a given number of points
in our treatment (points in our $\Theta$ dimension). This is primarily
due to the fact that \cite{shib_11, muel_10} use a variable Eddington
factor to close their system of equations. A comparison between this
method and a full Boltzmann equation, in our opinion, would be a very
valuable study. We discuss the collision operator in sparse detail,
and in future work we will investigate in greater detail this
operator. Particularly, it would be interesting to compare the
accuracy of the treatment with two moments of a Legendre decomposition
with our work.

Our work may be naturally extended by coupling it to a
(magneto)hydrodynamic code such as {\tt{CoCoNuT}}
\cite{dimm_05}. Before doing so, more work is needed to implement the
velocity dependent terms.

\section*{Acknowledgments}
We would like to thank the research group at l'Observatoire de Paris, Meudon
for their indispensable support and input, in particular \'Eric Gourgoulhon
and Philippe Grandcl\'ement for their insights and for proof reading our
manuscript. We would also like to thank Bernhard M\"uller, Thomas M\"adler,
and Isabel Cordero-Carri\'on for many valuable comments and insights,
and Nicolas Vasset for the calculations in \ref{app:transfomatrix}. This
work has been partially funded by the SN2NS project ANR-10-BLAN-0503.


\begin{appendix}
\section{Ricci rotation coefficients: additional terms}
\label{app:di0jdij0}

From equation~(\ref{eq:gammaij0}) we see that we need to calculate $d_{i 0 j}$
and $d_{i j 0}$, or equivalently, $d^i\,\!_{0 j}$ and $d^i\,\!_{j 0}$. The
algebraic steps needed to simplify these terms are presented in this appendix.

We first address the computation of the $d^i\,_{j 0}$ components.

With the use of equation~(\ref{eq:vierbderiv}) we obtain
\begin{equation}
  d^i\,\!_{j 0} = - M^{-1}\,\!\!_j\,\!^{\tilde{\mu}} \,
  M^{-1}\,\!\!_0\,\!^{\tilde{\nu}}  \, \frac{\partial 
  M^i\,\!_{\tilde{\mu}}}{\partial x^{\tilde{\nu}}}.
\end{equation}
Because
$M^{-1}\,\!\!_j\,\!^{0}=0$ in our $3+1$ formulation, $\tilde{\mu}$ may be
replaced by a spatial index, and thus
\begin{equation}
  d^i\,\!_{j 0} = - M^{-1}\,\!\!_j\,\!^{\tilde{k}} \,
  M^{-1}\,\!\!_0\,\!^{\tilde{\nu}} \, \frac{\partial 
  M^i\,\!_{\tilde{k}}}{\partial x^{\tilde{\nu}}}.
\end{equation}
Keeping in mind a $3+1$ approach we can then separate this into two terms:
the first corresponds to $\tilde{\nu}=0$ with $M^{-1}\,\!\!_0\,\!^0 = 1 /
\alpha$ and the second to $\tilde{\nu}=\tilde{l}=1,2,3$ with
$M^{-1}\,\!\!_0\,\!^{\tilde{l}} = -\beta^{\tilde{l}}/\alpha$,
\begin{equation}
  \label{eq:dij0terms}
  d^i\,\!_{j 0} = -\frac{1}{\alpha} M^{-1}\,\!\!_j\,\!^{\tilde{k}} \,
  \frac{\partial M^i\,\!_{\tilde{k}}}{\partial t} +
  M^{-1}\,\!\!_j\,\!^{\tilde{k}} \, \frac{\beta^{\tilde{l}}}{\alpha} \frac{\partial
    M^i\,\!_{\tilde{k}}}{\partial x^{\tilde{l}}} ~.
\end{equation}

With specialization to the CFC, the first term of
Eq.~(\ref{eq:dij0terms}) becomes
\begin{equation}
  -\frac{1}{\alpha} M^{-1}\,\!\!_j\,\!^{\tilde{k}} \,
  \frac{\partial M^i\,\!_{\tilde{k}}}{\partial t} = -\frac{1}{\alpha}
  \frac{1}{\Psi^2} \frac{\partial \Psi^2}{\partial t} \delta_j\,\!^i,
\end{equation}
where $\delta_j\,\!^i$ is the Kronecker delta. Using the maximal
slicing condition $K = K^i\,\!_i = 0$, we have for the conformal factor
\begin{equation}
  \frac{\partial \Psi}{\partial t} = \frac{\Psi}{6} \nabla_k \beta^k,
\end{equation}
(see \cite{cord_09}). We can then eliminate the explicit time derivative to get,
\begin{equation}
  -\frac{1}{\alpha} M^{-1}\,\!\!_j\,\!^{\tilde{k}} \,
  \frac{\partial M^i\,\!_{\tilde{k}}}{\partial t} = -\frac{1}{3\alpha}
  \nabla_k \beta^k \delta_j\,\!^i.
\end{equation}

Note that with another time slicing choice, or using another form for the spatial
metric, this derivation would have to be adapted. However, in principle, it is
still possible to get rid of the time derivative, following a similar
procedure.

We now explicit the second term of Eq.~(\ref{eq:dij0terms}). Within the CFC,
the purely spatial parts of $M$ and $M^{-1}$ are diagonal, so the only
non-zero spatial terms arise from $i=j = \tilde{k}$.

After some simplifications, we find:


\begin{eqnarray}
   M^{-1}\,\!\!_1\,\!^{\tilde{1}} \, \frac{\beta^{\tilde{l}}}{\alpha} \frac{\partial
    M^1\,\!_{\tilde{1}}}{\partial x^{\tilde{l}}} &=& \frac{1}{\alpha \Psi^2} \left[ \beta^r \frac{\partial
      \Psi^2}{\partial r} + \beta^\theta \frac{\partial
      \Psi^2}{\partial \theta} + \beta^\phi
    \frac{\partial \Psi^2}{\partial \phi} \right], \\
  M^{-1}\,\!\!_2\,\!^{\tilde{2}} \, \frac{\beta^{\tilde{l}}}{\alpha} \frac{\partial
    M^2\,\!_{\tilde{2}}}{\partial x^{\tilde{l}}} &=&\frac{1}{\alpha \Psi^2} \left[ \frac{\beta^r}{r} \frac{\partial
      \Psi^2 r}{\partial r} + \beta^\theta \frac{\partial
      \Psi^2}{\partial \theta} + \beta^\phi
    \frac{\partial \Psi^2}{\partial \phi} \right], \\
  M^{-1}\,\!\!_3\,\!^{\tilde{3}} \, \frac{\beta^{\tilde{l}}}{\alpha} \frac{\partial
    M^3\,\!_{\tilde{3}}}{\partial x^{\tilde{l}}} &=& 
  \frac{1}{\alpha \Psi^2} \left[ \frac{\beta^r}{r} \frac{\partial
      \Psi^2 r}{\partial r} +  \frac{\beta^\theta}{\sin \theta}
    \frac{\partial \Psi^2 \sin \theta}{\partial \theta} +
    \beta^\phi \frac{\partial \Psi^2}{\partial
      \phi} \right].
\end{eqnarray}
 
We now turn to the computation of the $d^i\,\!_{0 j}$ term. From
equation~(\ref{eq:vierbderiv}),
\begin{equation}
  d^i\,\!_{0 j} = -M^{-1}\,\!\!_0\,\!^{\tilde{\mu}} \,
  M^{-1}\,\!\!_j\,\!^{\tilde{\nu}} \,
  \frac{\partial M^i\,\!_{\tilde{\mu}}}{\partial x^{\tilde{\nu}}},
\end{equation}
where we see that $d^i\,\!_{0 j}$ is 0 if $\tilde{\nu}=0$ since
$M^{-1}\,\!\!_j\,\!^{0}=0$. Hence the fact that $\tilde{\nu}$ can be
replaced by a spatial index and $d^i\,\!_{0 j}$ does not contain any
time derivative. We also decompose $\tilde{\mu}$ in a $3+1$
fashion, with the temporal part and the spatial part,
\begin{equation}
  d^i\,\!_{0 j} = -M^{-1}\,\!\!_0\,\!^{0} \, M^{-1}\,\!\!_j\,\!^{\tilde{k}} \,
  \frac{\partial M^i\,\!_{0}}{\partial x^{\tilde{k}}} -
  M^{-1}\,\!\!_0\,\!^{\tilde{l}} \, M^{-1}\,\!\!_j\,\!^{\tilde{k}} \,
  \frac{\partial M^i\,\!_{\tilde{l}}}{\partial x^{\tilde{k}}}.
\end{equation}
With $M^{-1}\,\!\!_0\,\!^{0} = 1 / \alpha$ and $M^{-1}\,\!\!_0\,\!^{\tilde{l}} =
-\beta^{\tilde{l}} / \alpha$, we find
\begin{equation}
  \label{eq:di0jterms}
  d^i\,\!_{0 j} = -\frac{1}{\alpha} M^{-1}\,\!\!_j\,\!^{\tilde{k}} \,
  \frac{\partial M^i\,\!_{0}}{\partial x^{\tilde{k}}} +
  \frac{\beta^{\tilde{l}}}{\alpha} M^{-1}\,\!\!_j\,\!^{\tilde{k}} \,
  \frac{\partial M^i\,\!_{\tilde{l}}}{\partial x^{\tilde{k}}}.
\end{equation}

Within the CFC, after some simplifications, the first term of
Eq.(\ref{eq:di0jterms}) becomes
%
\begin{equation}
-\dfrac{1}{\alpha} M^{-1}\,\!\!_j\,\!^{\tilde{k}} \,
  \dfrac{\partial M^i\,\!_{0}}{\partial x^{\tilde{k}}}
   = -\dfrac{1}{\alpha \Psi^2} \begin{pmatrix}[2]
 \dfrac{\partial \Psi^2 \beta^r}{\partial r}  & \dfrac{\partial \Psi^2 
   r \beta^\theta}{\partial r}  & \sin \theta \dfrac{\partial \Psi^2 r \beta^\phi}{\partial r}  \\ 
 \dfrac{1}{r} \dfrac{\partial \Psi^2 \beta^r}{\partial \theta}  &
  \dfrac{\partial \Psi^2 \beta^\theta}{\partial \theta}  &
  \dfrac{\partial \Psi^2 \sin \theta \beta^\phi}{\partial \theta}  \\
 \dfrac{1}{r \sin \theta}\dfrac{\partial \Psi^2 \beta^r}{\partial \phi} 
 & \dfrac{1}{\sin \theta}\dfrac{\partial \Psi^2 
   \beta^\theta}{\partial \phi}  & \dfrac{\partial \Psi^2 \beta^\phi}{\partial \phi}  \\
\end{pmatrix}~,
\end{equation}
with $j$ the row index and $i$ the column index.

Within the CFC, after some simplifications, the second term of
Eq.(\ref{eq:di0jterms}) becomes,
%
\begin{equation}
  \dfrac{\beta^{\tilde{l}}}{\alpha} M^{-1}\,\!\!_j\,\!^{\tilde{k}} \,
  \dfrac{\partial M^i\,\!_{\tilde{l}}}{\partial x^{\tilde{k}}} 
  = \dfrac{1}{\alpha \Psi^2} \begin{pmatrix}[2]
  \beta^r \dfrac{\partial \Psi^2}{\partial r}  & \beta^\theta
  \dfrac{\partial \Psi^2 r}{\partial r}  & \beta^\phi \sin \theta
  \dfrac{\partial \Psi^2 r}{\partial r}  \\
  \dfrac{\beta^r}{r} \dfrac{\partial \Psi^2}{\partial \theta}  &
  \beta^\theta \dfrac{\partial \Psi^2}{\partial \theta}  &
  \beta^\phi \dfrac{\partial \Psi^2  \sin
    \theta}{\partial \theta}  \\ 
  \dfrac{\beta^r}{r \sin \theta} \dfrac{\partial \Psi^2}{\partial \phi}  &
  \dfrac{\beta^\theta}{\sin \theta} \dfrac{\partial \Psi^2}{\partial
    \phi}  & \beta^\phi \dfrac{\partial \Psi^2}{\partial \phi} 
  \end{pmatrix}~,
\end{equation}
where, as before, $j$ is the row index and $i$ the column index.

\section{Implementation of the Boltzmann solver}
\label{app:implementation}

We here give some technical details about \texttt{Ghost} for the numerical
integration of the Boltzmann equation~(\ref{eq:boltzefcfc}), which is based on
the numerical library \texttt{Lorene}~\cite{lorene}. However, \texttt{Ghost}
includes some generalization of our methods to the 6-dimensional case, as
\texttt{Lorene} is bound to the 3-dimensional case. We first give some general
concepts about spectral methods in one dimension, then the treatment of the
6-dimensional case is shown, then we detail the techniques for handling
coordinate singularities appearing in the Liouville operator in doubly
spherical coordinates, and finally we present the time integration scheme,
together with the treatment of boundary conditions.

\subsection{Spectral representation}
\label{app:spectral}

In order to represent a function $f(x)$, the most straightforward method is
to use a finite set of its values on a pre-defined grid. An
alternative, that we use here, is to represent it through a finite set
of {\em coefficients\/} of the decomposition of the function on a
pre-defined spectral basis. The most straightforward example is the use of a
truncated Fourier series for a periodic function:
\begin{equation}
  \label{e:ex_fourier}
  f(x) \simeq \sum_{k=0}^{N} a_k\, \cos (kx) + \sum_{k=1}^{N} b_k\, \sin (kx) ,
\end{equation}
where $f$ can be represented, up to some truncation error, by the set
of $2N+1$ coefficients $\{ a_k, b_k \}$. It can be shown (see
\textit{e.g.}~\cite{boyd-01, hesthaven-07}) that for
{\em smooth functions} this truncation error decreases faster than any power
of $N$. In practice, it is most often exponentially decreasing. If the
function $f$ is only continuous and $k$-times differentiable, then the
convergence of the truncation error goes as $1 / N^{k+1}$. If $f$ is
not periodic or discontinuous, then there is no convergence of the
functional series~(\ref{e:ex_fourier}) to $f$; this is called the {\em
  Gibbs phenomenon}~\cite{boyd-01}. 

When dealing with a non-periodic function, one can replace the Fourier
basis by a set of orthogonal polynomials. The family of Jacobi
polynomials~\cite{hesthaven-07} is well-suited for such
representations with the particular cases of Legendre and Chebyshev
polynomials. In our work, we use the Chebyshev polynomials to describe
the non-periodic variables of our distribution function. In the case
of a function $g(y)$ depending on the single variable $y\in [-1,1]$:
\begin{equation}
  \label{e:ex_cheb}
  g(y) \simeq \sum_{k=0}^N c_k \, T_k(y), \textrm{ where } T_k(y) =
  \cos \left( k \arccos (y) \right)
\end{equation}
is the $k$-th Chebyshev polynomial. Convergence rates of this series
are the same as for Fourier series, depending on the degree of
differentiability of $g(y)$. In \ref{app:6D_spectral}, we shall
detail the implementation of the spectral representation of the
distribution depending on six variables. We here first give the
methods used in one dimension, using Fourier or Chebyshev
decomposition basis, to compute derivatives or integrals.

If the Fourier coefficients of a periodic function $f(x)$ are known,
it is rather simple to determine those of its derivative:
\begin{equation}
  \label{e:ex_fourier_der}
  f'(x) = \frac{\textrm{d}f}{\textrm{d}x} \simeq \sum_{k=1}^N
  \tilde{a}_k\, \cos (k x) + \tilde{b}_k\, \sin (k x),
\end{equation}
where we can deduce the the $\{ \tilde{a}_k, \tilde{b}_k \}$ from the
$\{ a_k, b_k \}$, through
\begin{equation}
  \label{e:derivee_fourier}
  \tilde{a}_k = k\, b_k, \textrm{ and } \tilde{b}_k = -k\, a_k.
\end{equation}
With Chebyshev polynomials, the coefficients $\tilde{c}_k$ of the
derivative $g'(y)$ are given by~\cite{hesthaven-07}:
\begin{equation}
  \label{e:derivee_cheb}
  \tilde{c}_k = \frac{2}{1 + \delta_0^{\ k}} \sum_{\substack{j=k+1\\j+k \textrm{
      odd}}}^N j\, c_j.
\end{equation}
The computation of integral terms is done very easily with Fourier
series: if the integral is taken over a period, one only needs to
consider the coefficient $a_0$, the others terms giving
zero. Integration of Chebyshev polynomials over the interval $[-1,1]$ gives:
\begin{equation}
  \label{e:integ_cheb}
  \int_{-1}^1 T_k(y)\, \textrm{d}y = \left\{
    \begin{array}{cl}
      -\dfrac{2}{k^2-1}& k \textrm{ even},\\
      0 & k \textrm{ odd}.
    \end{array}
  \right.
\end{equation}

Spectral coefficients $\{ a_k, b_k \}$ or $\{c_k\}$ can be computed
\textit{via} quadrature formula
(see \textit{e.g.}~\cite{hesthaven-07}), or using a Fast Fourier
  Transform (FFT). Indeed, from the definition of Chebyshev
polynomials~(\ref{e:ex_cheb}), it is possible to use a FFT to compute
the coefficients $\{c_k\}$~\cite{boyd-01}. Any of these methods
require the knowledge of the values of function on a numerical grid:
either equally spaced points for Fourier decomposition, or collocation
points $\{ x_k = -\cos\left( k\pi / N \right) \}$ for Chebyshev
polynomials. Moreover, when non-linear terms are to be evaluated, as
the product of two functions, the easiest way is to compute this
product on the numerical grid, without using the coefficients (which
would require the calculation of a convolution sum). Therefore, when
representing a function $f(x)$ we will use a dual form with both
descriptions: the values at grid points $\{ f( x_k ) \}$ and the
coefficients $\{ a_k, b_k \}$ or $\{c_k\}$, with the use of FFT to
update each part, when needed. In what follows, we write the Fourier
coefficients globally as
\begin{equation}
  \label{e:one_coef}
  f(x) \simeq \sum_{k=0}^n c_k \, F_k(x) = \sum_{k=0}^{n/2} a_k\, \cos
  (kx) + \sum_{k=1}^{n/2} b_k\, \sin (kx). 
\end{equation}

\subsection{Six-dimensional representation of the distribution
  function}
\label{app:6D_spectral}

At a given time-step $t$, the distribution function $f_t$ depends
on six spatial variables in doubly spherical coordinates, and can be
decomposed onto Chebyshev bases for the $r, \epsilon$ and $\Theta$
variables, whereas Fourier bases are used for $\theta\footnote{The
  treatment of the $\theta$ argument is based on an analytic extension
  of the represented function to the whole interval $\left[ 0, 2\pi
  \right)$, that makes it periodic in $\theta$ and thus convenient for
  a Fourier representation.}, \phi$ and $\Phi$ variables. This can be
written as follows:
\begin{equation}
  \label{e:6_variables}
  f_t \left(r, \theta, \phi, \epsilon, \Theta, \Phi \right) \simeq
\sum_{i=0}^{n_r} \sum_{j=0}^{n_\theta} \sum_{k=0}^{n_\phi} \sum_{l=0}^{n_\epsilon}
\sum_{m=0}^{n_\Theta} \sum_{i=0}^{n_\Phi} C_{ijklmp}\, T_i (\bar{r})\,
F_j(\theta)\, F_k(\phi)\, T_l( \bar{\epsilon} )\, T_m( \bar{\Theta}
)\, F_p(\Phi).
\end{equation}
The variables with a bar above belong to the interval $[-1,1]$ and are
related to the coordinate variables through simple affine change,
\textit{e.g.}:
\begin{equation}
  \label{e:map_af}
  r = \alpha_r \bar{r} + \beta_r, \qquad \bar{r} \in [-1,1],
\end{equation}
where $(\alpha_r, \beta_r)$ are two constants, such that $R_{\rm min}
= \beta_r - \alpha_r$ and $R_{\rm max} = \alpha_r + \beta_r$. A similar
setting is chosen for the $\epsilon$ variable.

The nature of the Boltzmann equation explicitly requires a different
treatment of boundary condition whether $\Theta \geq \pi/2$ or not.
Indeed, $\Theta$ physically represents the direction (outgoing if $0
\le \Theta \le \pi/2$ or ingoing if $\pi/2 \le \Theta \le \pi$) of the
neutrinos. For instance, the surface of an isolated proto-neutron star
has an outgoing flux but no incoming flux, the distribution function
is expected to be $\mathcal{C}^0$, but not differentiable, at $\Theta
= \pi/2$ (see \textit{e.g.}~\cite{bona_11}). In a core-collapse
context, the proto-neutron star is not isolated but can approach the
described case enough to justify our special treatment of the
$\Theta=\pi/2$ point. This is described in detail in
\cite{bona_11}. As that path-finding study described, we also use the
Chebyshev basis for the $\Theta$ argument, within the context of a
multi-domain approach~\cite{canuto-07}. In particular, we split our
$\Theta$ interval into two domains, separated at $\Theta=\pi/2$, as
this allows us to benefit from the fast convergence of spectral
methods despite the fact that the solution is only $\mathcal{C}^0$
continuous across this domain boundary. An affine change of variables
of the type~(\ref{e:map_af}) is defined in each domain.

Particular care must be taken when dealing with the polar coordinate
$\theta$. We have modified the existing methods in the {\tt{Lorene}}
library for this coordinate to allow for dependence on both the
azimuthal ($\phi$) and momentum azimuthal ($\Phi$) coordinates. Based
on regularity properties, the $\theta$ representation
($\sin{(j\theta)}$ or $\cos{(j\theta)}$) is deduced from the $\phi$
representation ($\sin{(m\phi)}$ or $\cos{(m\phi)}$) and the $\Phi$
representation ($\sin{(M\Phi)}$ or $\cos{(M\Phi)}$).

Let us, for a moment, consider a 3D scalar field $F (r, \theta,
\phi)$. The representation of $F$ with {\tt{Lorene}} has already been
described \textit{e.g.} in~\cite{gran_09}, and our treatment of a distribution
function (a 6D scalar field) is a natural generalization of it. The
technical point here is that the exact $\theta$ basis in
{\tt{Lorene}} is dependent on the nature of the $\phi$ and $\Phi$
bases, due to the singular character of spherical coordinates and to
the regularity requirement on the axis $\theta=0$ of our functions
(see~\cite{gran_09} for details).

Typically, we expect that the decomposition~(\ref{e:6_variables}) for
functions $F(\theta,\phi)$ may be written as a linear combination of,
\begin{equation}\label{eq:3DBasis}
 \sin{(j\theta)}\sin{(m\phi)},\,
 \sin{(j\theta)}\cos{(m\phi)},\,
 \cos{(j\theta)}\sin{(m\phi)},\,
 \cos{(j\theta)}\cos{(m\phi)}
\end{equation}
The $\theta$ argument is decomposed on a basis with sines and
cosines, depending on the parity of $m$, {\tt{COSSIN\_C}} or
{\tt{COSSIN\_S}} which are defined as follows.

For regular scalar fields in 3D case, one needs the {\tt{COSSIN\_C}}
basis, which  is defined to be a series that uses the base elements
\begin{equation}
\left.
\begin{array}{c}
 \cos{\left(j\theta\right)}\sin{(m\phi)}\\
 \cos{\left(j\theta\right)}\cos{(m\phi)}
\end{array}\right\} \rm{m\, even},
\end{equation}
and
\begin{equation}
\left.
\begin{array}{c}
 \sin{\left(j\theta\right)}\sin{(m\phi)}\\
 \sin{\left(j\theta\right)}\cos{(m\phi)}
\end{array}\right\} \rm{m\, odd},
\end{equation}
In addition, the {\tt{COSSIN\_S}} basis uses,
\begin{equation}
\left.
\begin{array}{c}
 \sin{\left(j\theta\right)}\sin{(m\phi)}\\
 \sin{\left(j\theta\right)}\cos{(m\phi)}
\end{array}\right\} \rm{m\, even},
\end{equation}
and
\begin{equation}
\left.
\begin{array}{c}
 \cos{\left(j\theta\right)}\sin{(m\phi)}\\
 \cos{\left(j\theta\right)}\cos{(m\phi)}
\end{array}\right\} \rm{m\, odd}.
\end{equation}

So, {\tt{COSSIN}} means an expansion in both cosines and sines for
$\theta$, and {\tt{\_C}} (for cosine) or {\tt{\_S}} (for sine) refers
to the expansion corresponding to $\cos{(0 \, \phi)}$. It is
straightforward to see that some operators, for instance
$\partial / \partial \theta$, switch the basis from {\tt{COSSIN\_C}} to
{\tt{COSSIN\_S}} and \textit{vice-versa}. This is the reason why both
bases are implemented.

Coming back to the six dimensional calculation, we consider
expanding this definition to include the Fourier basis for the $\Phi$
argument. The base elements for a function $G(\theta,\phi,\Phi)$ are,
\begin{equation}
\begin{matrix}
 \sin{(j\theta)}\sin{(m\phi)}\sin{(M\Phi)},\quad
 \sin{(j\theta)}\sin{(m\phi)}\cos{(M\Phi)},\\
 \cos{(j\theta)}\sin{(m\phi)}\sin{(M\Phi)},\quad
 \cos{(j\theta)}\sin{(m\phi)}\cos{(M\Phi)},\\
 \sin{(j\theta)}\cos{(m\phi)}\sin{(M\Phi)},\quad
 \sin{(j\theta)}\cos{(m\phi)}\cos{(M\Phi)},\\
 \cos{(j\theta)}\cos{(m\phi)}\sin{(M\Phi)},\quad
 \cos{(j\theta)}\cos{(m\phi)}\cos{(M\Phi)}.
\end{matrix}
\end{equation}
Analogously, the expansion rule can be inferred from the 3D treatment,
but instead of considering the parity of $m$ alone, we consider the
parity of the combination $m+M$. While this approach may appear to be
complex, it allows us to manipulate a (regular) six dimensional scalar
field in a fully consistent manner.

In the 6D case, for the decomposition of the distribution function,
the new {\tt{COSSIN\_C}} basis corresponds to
\begin{equation}
\left.
\begin{array}{c}
 \cos{\left(j\theta\right)}\cos{(m\phi)}\sin{(M\Phi)}\\
 \cos{\left(j\theta\right)}\sin{(m\phi)}\sin{(M\Phi)}\\
 \cos{\left(j\theta\right)}\cos{(m\phi)}\cos{(M\Phi)}\\
 \cos{\left(j\theta\right)}\sin{(m\phi)}\cos{(M\Phi)}
\end{array}\right\} \rm{(M+m)\, even},
\end{equation}
and
\begin{equation}
\left.
\begin{array}{c}
 \sin{\left(j\theta\right)}\cos{(m\phi)}\sin{(M\Phi)}\\
 \sin{\left(j\theta\right)}\sin{(m\phi)}\sin{(M\Phi)}\\
 \sin{\left(j\theta\right)}\cos{(m\phi)}\cos{(M\Phi)}\\
 \sin{\left(j\theta\right)}\sin{(m\phi)}\cos{(M\Phi)}
\end{array}\right\} \rm{(M+m)\, odd}
\end{equation}
While the new {\tt{COSSIN\_S}} basis uses
\begin{equation}
\left.
\begin{array}{c}
 \cos{\left(j\theta\right)}\cos{(m\phi)}\sin{(M\Phi)}\\
 \cos{\left(j\theta\right)}\sin{(m\phi)}\sin{(M\Phi)}\\
 \cos{\left(j\theta\right)}\cos{(m\phi)}\cos{(M\Phi)}\\
 \cos{\left(j\theta\right)}\sin{(m\phi)}\cos{(M\Phi)}
\end{array}\right\} \rm{(M+m)\, odd},
\end{equation}
and
\begin{equation}
\left.
\begin{array}{c}
 \sin{\left(j\theta\right)}\cos{(m\phi)}\sin{(M\Phi)}\\
 \sin{\left(j\theta\right)}\sin{(m\phi)}\sin{(M\Phi)}\\
 \sin{\left(j\theta\right)}\cos{(m\phi)}\cos{(M\Phi)}\\
 \sin{\left(j\theta\right)}\sin{(m\phi)}\cos{(M\Phi)}
\end{array}\right\} \rm{(M+m)\, even}.
\end{equation}

\subsection{Singularity-avoiding technique}

\label{app:singularities}
There are two kinds of coordinate singularities which arise in our
formulation. For instance, in Eq.~(\ref{eq:fullspatialnonrel}) one can
see that a coordinate singularity occurs at the origin where the
radial coordinate $r$ goes to $0$, with the term $r^{-1}$ in front of
all angular derivatives (in a similar way as in many other operators, like the
Laplace operator). Another coordinate singularity occurs on the polar axis
(with $\sin^{-1}(\theta)$ terms), again, akin to the Laplace operator this
occurs on the terms with azimuthal and momentum-azimuthal derivatives in
Eq.~(\ref{eq:fullspatialnonrel}). Momentum-polar axis singularity (with
$\sin^{-1}(\Theta)$ terms) appears in the relativistic Liouville operator, as
can be seen in the expression of the Jacobian
Eq.~(\ref{eq:momspcjacobian}). Additional $\sin^{-1}(\theta)$ terms also occur
in the Ricci rotation coefficients.

The singularity at the origin is not treated in this paper. The long
term strategy is to use a simplified version in the innermost part,
and to couple it to our full Boltzmann equation, that would take care
of all other parts. The simplified version could be a diffusion
equation or a telegraph equation (see \cite{bona_11} for a detailed
discussion and implementation of the telegraph equation, as well as
for the matching between both equations). This treatment is known to
be sufficient in the innermost part of a core-collapse supernova
simulation, and there are well known techniques to handle the
singularity at the origin when using these approaches, described in
\cite{gran_09}. In this work, we restrict ourselves to radial shells
from a given $R_{\rm min}>0$ to a given $R_{\rm max}$.

The axial singularity is treated in the same fashion as in {\tt
  Lorene} in 3D. {\tt Lorene}'s operators are expressed using
spherical geometry. If we take the Laplace operator as an example, we
know that the full Laplace operator is regular, despite the individual terms
being singular, if expressed in spherical coordinates. When combined,
the apparent singularities parts analytically cancel and thus we do
not need to compute them. Instead, singular operator as $\sin^{-1}
(\theta)$ are computed only through their finite part (see
also~\cite{gran_09}). We exploit the same trends in our numerical
development of the Liouville operator.

Practically speaking, when coding our solution we use the computed
quantity $\bar{\Gamma}^i\,_{\mu \nu} \dfrac{\partial f}{\partial
  p^i}$. We first compute the partial derivative with respect to
$p^i$, and then multiply by each term of the corresponding connection
coefficient. The multiplication by the terms that may contain
$\sin^{-1} (\theta)$ corresponds to the last step of the computation.

\subsection{Time integration}
\label{app:time}

With the tools described in \ref{app:spectral}, \ref{app:6D_spectral}
and \ref{app:singularities} hereabove, it is possible to compute the
spatial part of the relativistic Liouville operator, acting on a
distribution function, as well as the collision
term~(\ref{e:collision}).

We note here $\tilde{L}$ the spatial part of the relativistic differential
Liouville operator $L[f]$~(\ref{eq:boltzefcfc}):
\begin{equation}
  \label{e:spatial_Liouville}
  \tilde{L}[f] =  \alpha \left[ \left(
    \frac{p^i}{\Psi^2 \epsilon} -
  \frac{\beta^i}{\alpha} \right) \frac{\partial f}{\partial x^i} -
\bar{\Gamma}^j\,\!_{\mu \nu} p^\mu p^\nu J^i\,\!_j
\frac{1}{\epsilon}\frac{\partial f}{\partial p^i} \right].
\end{equation}
The time integration of the relativistic Boltzmann equation for the
distribution function $f(t, r, \theta, \phi, \epsilon, \Theta, \Phi)$,
without the collision term can thus be written as
\begin{equation}
  \label{e:time_pde}
  \frac{\partial f}{\partial t} = - \tilde{L}[f]. 
\end{equation}

At every time $t$, the spectral approximation to the distribution
function $f(t, \dots)$ is written as $U_N(t)$, the finite set of
total size $N= n_r \times n_\theta \times n_\phi \times n_\epsilon
\times n_\Theta \times n_\Phi$, composed of the time-dependent
spectral coefficients $C_{ijklmp}$ of
Eq.~(\ref{e:6_variables}). We note $\tilde{L}_N$ the spectral
approximation to the operator $\tilde{L}$, together with the boundary
conditions which will be detailed hereafter. $\tilde{L}_N$ can be
represented by an $N\times N$ matrix, although this matrix is never
computed explicitly in our code. With this approach, we can use the
so-called method of lines, which allows one to reduce a PDE to an ODE,
after discretization in all but one dimensions. We can then, in
principle, use any of the well-known ODE integration schemes (see
\textit{e.g.}~\cite{hesthaven-07}). 

In \texttt{Ghost} we have, for the moment, implemented only
equally-spaced grid in time, with the third-order Adams-Bashforth
scheme:
\begin{equation}
  \label{e:ODE_integration}
  U_N^{J+1} = U_N^J - \Delta t \left( \frac{23}{12} \tilde{L}_N U_N^J
    - \frac{4}{3} \tilde{L}_N U_N^{J-1} + \frac{5}{12} \tilde{L}_N
    U_N^{J-2} \right) ,
\end{equation}
with $U_N^J = U_N(t = J\times \Delta t)$, $\Delta t$ being the
time-step. This is an explicit scheme but, in principle, implicit or
semi-implicit (see also~\cite{gran_09}) schemes could be implemented. 
Thanks to the method of lines, usual stability analysis can be applied
to any of the time-marching scheme, as $\tilde{L}_N$ can be
diagonalized and it is possible to study the collection of scalar ODE
problems: 
\begin{equation}
  \label{e:eigen_dt}
  \frac{\partial U_N}{\partial t} = \lambda_i U_N,
\end{equation}
where $\lambda_i$ is any of the eigenvalues of $\tilde{L}_N$. Standard
result is that, of course, explicit schemes as the one we use here,
are subject to the Courant-Friedrichs-Lewy (CFL) stability condition:
the maximum time-step is of the order of magnitude of the time scale
for propagation across the smallest distance between two grid
points~\cite{boyd-01}, \cite{gran_09}. From the definition of the
grids we use in our code (see \ref{app:spectral} above), this distance
goes as $1/n$ for the Fourier variables ($\theta, \phi, \Phi$) and
$1/n^2$ for the Chebyshev ones ($r, \epsilon, \Theta$).

Boundary conditions are imposed using the {\em tau method}, which
consists in modifying the last spectral coefficient, for each variable
with a boundary condition, at each time-step. To illustrate this
point, let us consider the simple advection PDE for an unknown
function $v(t, x)$:
\begin{eqnarray}
  \label{e:simple_PDE}
  \forall t\geq 0, \forall x \in [-1,1], \quad &&\frac{\partial
    v}{\partial t } = \frac{\partial v}{\partial x} \\
  \forall t\geq 0, \quad &&v(t, 1) = b(t), \nonumber
\end{eqnarray}
with $b(t)$ a given function. Let $V_N^J$ be the vector composed of
the $N+1$ coefficients of $v(t=J\times \Delta t, x)$ in the Chebyshev
basis. If we denote by $\ell_N$ the spectral approximation of the
operator $\partial / \partial x$ and $\left( \ell_N V_N^J \right)_i$
the $i$-th coefficient of $\ell_N$ applied to $V_N^J$, then one can advance
in time using the third-order Adams-Bashforth scheme:
\begin{eqnarray}
   V_i^{J+1} = V_i^J + \Delta t \left(
    \frac{23}{12} \ell_N V_N^J - \frac{4}{3} \ell_N V_N^{J-1} +
    \frac{5}{12} \ell_N V_N^{J-2} \right)_i \quad i=0\dots
  N-1, \nonumber \\
  V_N^{J+1} =b((J+1)\Delta t) - \sum_{k=0}^{N-1} V_k^{J+1} \label{e:syst_tau}
 .
\end{eqnarray}
The expression on the last line comes from the fact that $T_k(x=1) =
1$. Boundary conditions are thus imposed at $r=R_{\rm min}$ for
$0 \leq \Theta \leq \pi/2$ and $r=R_{\rm max}$ for
$\pi/2 \leq \Theta \leq \pi$, following~\cite{bona_11}.

\section{Doubly spherical coordinates transformation matrix $P$}
\label{app:transfomatrix}

In the $\vec{e}_{r}, \vec{e}_{\theta}, \vec{e}_{\phi}$ basis,
$\vec{p}$ is defined by Eq.(\ref{eq:pcartreal}). We then apply two
rotations of angles $\theta$ and $\phi$ to express $\vec{p}$ in
the $\vec{e}_{x}, \vec{e}_{y}, \vec{e}_{z}$ basis (this is the usual
transformation from a spherical to a Cartesian basis),

\begin{eqnarray}
  p_x &=& \epsilon (\cos \Theta \sin \theta \cos \phi + \sin
  \Theta \cos \Phi \cos \theta \cos \phi - \sin \Theta \sin \Phi
  \sin \phi)~,\nonumber\\
  p_y &=& \epsilon (\cos \Theta \sin \theta \sin \phi + \sin \Theta
  \cos \Phi \cos \theta \sin \phi + \sin \Theta \sin \Phi \cos
  \phi)~,\nonumber\\
  p_z &=& \epsilon (\cos \Theta \cos \theta - \sin \Theta \cos \Phi
  \sin \theta)~.
\end{eqnarray}

We then compute the transformation matrix $P_1$,
\begin{equation}
P_1 = \dfrac{\partial (p_x, p_y, p_z)}{\partial (\epsilon, \Theta,
  \Phi)}~,
\end{equation}
with
\begin{eqnarray}
  \frac{\partial p_x}{\partial \epsilon} &=& \frac{p_x}{\epsilon},
  \quad \quad \frac{\partial p_y}{\partial \epsilon} =
  \frac{p_y}{\epsilon}, \quad \quad \frac{\partial p_z}{\partial
    \epsilon} = \frac{p_z}{\epsilon},\nonumber\\
  \frac{\partial p_x}{\partial \Theta} &=& \epsilon (-\sin \Theta \sin
  \theta \cos \phi + \cos \Theta \cos \Phi \cos \theta \cos \phi -
  \cos \Theta \sin \Phi \sin \phi), \nonumber\\
  \frac{\partial p_y}{\partial \Theta} &=& \epsilon (-\sin \Theta \sin
  \theta \sin \phi + \cos \Theta \cos \Phi \cos \theta \sin \phi +
  \cos \Theta \sin \Phi \cos \phi), \nonumber\\
  \frac{\partial p_z}{\partial \Theta} &=& \epsilon (-\sin \Theta \cos
  \theta - \cos \Theta \cos \Phi \sin \theta), \nonumber\\
  \frac{\partial p_x}{\partial \Phi} &=& \epsilon (-\sin \Theta \sin
  \Phi \cos \theta \cos \phi - \sin \Theta \cos \Phi \sin \phi), \nonumber\\
  \frac{\partial p_y}{\partial \Phi} &=& \epsilon (-\sin \Theta \sin
  \Phi \cos \theta \sin \phi + \sin \Theta \cos \Phi \cos \phi), \nonumber\\
  \frac{\partial p_z}{\partial \Phi} &=& \epsilon (\sin \Theta \sin
  \Phi \sin \theta). \nonumber
\end{eqnarray}
We then compute the transformation matrix $P_2$,
\begin{equation}
P_2 = \dfrac{\partial (p_x, p_y, p_z)}{\partial (r, \theta, \phi)}~,
\end{equation}
with
\begin{eqnarray}
  \frac{\partial p_x}{\partial r} &=& 0, \quad \quad \frac{\partial
    p_y}{\partial r} = 0, \quad \quad \frac{\partial p_z}{\partial r}
  = 0,\nonumber\\ 
  \frac{\partial p_x}{\partial \theta} &=& \epsilon (\cos \Theta \cos
  \theta \cos \phi - \sin \Theta \cos \Phi \sin \theta \cos \phi), \nonumber\\
  \frac{\partial p_y}{\partial \theta} &=& \epsilon (\cos \Theta \cos
  \theta \sin \phi - \sin \Theta \cos \Phi \sin \theta \sin \phi), \nonumber\\
  \frac{\partial p_z}{\partial \theta} &=& \epsilon (-\cos \Theta \sin
  \theta - \sin \Theta \cos \Phi \cos \theta), \nonumber\\
  \frac{\partial p_x}{\partial \phi} &=& \epsilon (-\cos \Theta \sin
  \theta \sin \phi - \sin \Theta \cos \Phi \cos \theta \sin \phi -
  \sin \Theta \sin \Phi \cos \phi), \nonumber\\
  \frac{\partial p_y}{\partial \phi} &=& \epsilon (\cos \Theta \sin
  \theta \cos \phi + \sin \Theta \cos \Phi \cos \theta \cos \phi -
  \sin \Theta \sin \Phi \sin \phi), \nonumber\\
  \frac{\partial p_z}{\partial \phi} &=& 0. \nonumber
\end{eqnarray}
The transformation matrix we need, $P$, is then
\begin{displaymath}
  P = P_1^{-1} \times P_2~,
\end{displaymath}
\begin{equation}
  P = \begin{pmatrix}[2]
0, & 0, & 0, \\
0, & \cos \Phi, & \sin \Phi \sin \theta, \\
0, & - \dfrac{\sin \Phi \cos \Theta}{\sin \Theta}, & \cos \theta + \sin
\theta \dfrac{\cos \Phi \cos \Theta}{\sin \Theta},
\end{pmatrix}~,
\end{equation}
which is the one used in section~\ref{ss:nonrel_liouville}.

\end{appendix}
\section*{Bibliography}
\bibliographystyle{unsrt}

\input{articleboltz_gr.bbl}

\end{document}

%% file: sph6d.eps_tex
\begingroup%
  \makeatletter%
  \providecommand\color[2][]{%
    \errmessage{(Inkscape) Color is used for the text in Inkscape, but the package 'color.sty' is not loaded}%
    \renewcommand\color[2][]{}%
  }%
  \providecommand\transparent[1]{%
    \errmessage{(Inkscape) Transparency is used (non-zero) for the text in Inkscape, but the package 'transparent.sty' is not loaded}%
    \renewcommand\transparent[1]{}%
  }%
  \providecommand\rotatebox[2]{#2}%
  \ifx\svgwidth\undefined%
    \setlength{\unitlength}{349bp}%
    \ifx\svgscale\undefined%
      \relax%
    \else%
      \setlength{\unitlength}{\unitlength * \real{\svgscale}}%
    \fi%
  \else%
    \setlength{\unitlength}{\svgwidth}%
  \fi%
  \global\let\svgwidth\undefined%
  \global\let\svgscale\undefined%
  \makeatother%
  \begin{picture}(1,1.13669863)%
    \put(0,0){\includegraphics[width=\unitlength]{sph6d.eps}}%
    \put(0.73890192,0.91683593){\color[rgb]{0,0,0}\makebox(0,0)[lb]{\smash{$\Theta$}}}%
    \put(0.65429141,0.88357824){\color[rgb]{0,0,0}\makebox(0,0)[lb]{\smash{$\Phi$}}}%
    \put(0.02900107,0.0020214){\color[rgb]{0,0,0}\makebox(0,0)[lb]{\smash{$x$}}}%
    \put(0.29893516,0.24462043){\color[rgb]{0,0,0}\makebox(0,0)[lb]{\smash{$\phi$}}}%
    \put(0.32797869,0.54786914){\color[rgb]{0,0,0}\makebox(0,0)[lb]{\smash{$\theta$}}}%
    \put(0.5398257,0.60681048){\color[rgb]{0,0,0}\makebox(0,0)[lb]{\smash{$r$}}}%
    \put(0.96098935,0.38642123){\color[rgb]{0,0,0}\makebox(0,0)[lb]{\smash{$y$}}}%
    \put(0.30238436,1.07919513){\color[rgb]{0,0,0}\makebox(0,0)[lb]{\smash{$z$}}}%
    \put(0.82603249,0.85447767){\color[rgb]{0,0,0}\makebox(0,0)[lb]{\smash{$\vec{e}_r$}}}%
    \put(0.84954021,0.76099009){\color[rgb]{0,0,0}\makebox(0,0)[lb]{\smash{$\vec{e}_\phi$}}}%
    \put(0.79230733,0.69094385){\color[rgb]{0,0,0}\makebox(0,0)[lb]{\smash{$\vec{e}_\theta$}}}%
    \put(0.59156519,1.11549216){\color[rgb]{0,0,0}\makebox(0,0)[lb]{\smash{$\vec{p}$}}}%
    \put(0.67,0.79){\color[rgb]{0,0,0}\makebox(0,0)[lb]{\smash{$M$}}}%
    \put(0.7,0.05){\color[rgb]{0,0,0}\makebox(0,0)[lb]{\smash{$M'$}}}%
  \end{picture}%
\endgroup%

%% file: new_conserv_relat_15.tex
\begingroup
  \makeatletter
  \providecommand\color[2][]{%
    \GenericError{(gnuplot) \space\space\space\@spaces}{%
      Package color not loaded in conjunction with
      terminal option `colourtext'%
    }{See the gnuplot documentation for explanation.%
    }{Either use 'blacktext' in gnuplot or load the package
      color.sty in LaTeX.}%
    \renewcommand\color[2][]{}%
  }%
  \providecommand\includegraphics[2][]{%
    \GenericError{(gnuplot) \space\space\space\@spaces}{%
      Package graphicx or graphics not loaded%
    }{See the gnuplot documentation for explanation.%
    }{The gnuplot epslatex terminal needs graphicx.sty or graphics.sty.}%
    \renewcommand\includegraphics[2][]{}%
  }%
  \providecommand\rotatebox[2]{#2}%
  \@ifundefined{ifGPcolor}{%
    \newif\ifGPcolor
    \GPcolortrue
  }{}%
  \@ifundefined{ifGPblacktext}{%
    \newif\ifGPblacktext
    \GPblacktexttrue
  }{}%
  \let\gplgaddtomacro\g@addto@macro
  \gdef\gplbacktext{}%
  \gdef\gplfronttext{}%
  \makeatother
  \ifGPblacktext
    \def\colorrgb#1{}%
    \def\colorgray#1{}%
  \else
    \ifGPcolor
      \def\colorrgb#1{\color[rgb]{#1}}%
      \def\colorgray#1{\color[gray]{#1}}%
      \expandafter\def\csname LTw\endcsname{\color{white}}%
      \expandafter\def\csname LTb\endcsname{\color{black}}%
      \expandafter\def\csname LTa\endcsname{\color{black}}%
      \expandafter\def\csname LT0\endcsname{\color[rgb]{1,0,0}}%
      \expandafter\def\csname LT1\endcsname{\color[rgb]{0,1,0}}%
      \expandafter\def\csname LT2\endcsname{\color[rgb]{0,0,1}}%
      \expandafter\def\csname LT3\endcsname{\color[rgb]{1,0,1}}%
      \expandafter\def\csname LT4\endcsname{\color[rgb]{0,1,1}}%
      \expandafter\def\csname LT5\endcsname{\color[rgb]{1,1,0}}%
      \expandafter\def\csname LT6\endcsname{\color[rgb]{0,0,0}}%
      \expandafter\def\csname LT7\endcsname{\color[rgb]{1,0.3,0}}%
      \expandafter\def\csname LT8\endcsname{\color[rgb]{0.5,0.5,0.5}}%
    \else
      \def\colorrgb#1{\color{black}}%
      \def\colorgray#1{\color[gray]{#1}}%
      \expandafter\def\csname LTw\endcsname{\color{white}}%
      \expandafter\def\csname LTb\endcsname{\color{black}}%
      \expandafter\def\csname LTa\endcsname{\color{black}}%
      \expandafter\def\csname LT0\endcsname{\color{black}}%
      \expandafter\def\csname LT1\endcsname{\color{black}}%
      \expandafter\def\csname LT2\endcsname{\color{black}}%
      \expandafter\def\csname LT3\endcsname{\color{black}}%
      \expandafter\def\csname LT4\endcsname{\color{black}}%
      \expandafter\def\csname LT5\endcsname{\color{black}}%
      \expandafter\def\csname LT6\endcsname{\color{black}}%
      \expandafter\def\csname LT7\endcsname{\color{black}}%
      \expandafter\def\csname LT8\endcsname{\color{black}}%
    \fi
  \fi
  \setlength{\unitlength}{0.0500bp}%
  \begin{picture}(7200.00,5040.00)%
    \gplgaddtomacro\gplbacktext{%
      \csname LTb\endcsname%
      \put(1320,704){\makebox(0,0)[r]{\strut{} $-4\times 10^{-5}$}}%
      \put(1320,1111){\makebox(0,0)[r]{\strut{} $-2\times 10^{-5}$}}%
      \put(1320,1518){\makebox(0,0)[r]{\strut{} 0}}%
      \put(1320,1925){\makebox(0,0)[r]{\strut{} $2\times 10^{-5}$}}%
      \put(1320,2332){\makebox(0,0)[r]{\strut{} $4\times 10^{-5}$}}%
      \put(1320,2740){\makebox(0,0)[r]{\strut{} $6\times 10^{-5}$}}%
      \put(1320,3147){\makebox(0,0)[r]{\strut{} $8\times 10^{-5}$}}%
      \put(1320,3554){\makebox(0,0)[r]{\strut{} $10\times 10^{-5}$}}%
      \put(1320,3961){\makebox(0,0)[r]{\strut{} $12\times 10^{-5}$}}%
      \put(1320,4368){\makebox(0,0)[r]{\strut{} $14\times 10^{-5}$}}%
      \put(1320,4775){\makebox(0,0)[r]{\strut{} $16\times 10^{-5}$}}%
      \put(1452,484){\makebox(0,0){\strut{} 0}}%
      \put(1987,484){\makebox(0,0){\strut{} 5}}%
      \put(2522,484){\makebox(0,0){\strut{} 10}}%
      \put(3057,484){\makebox(0,0){\strut{} 15}}%
      \put(3592,484){\makebox(0,0){\strut{} 20}}%
      \put(4128,484){\makebox(0,0){\strut{} 25}}%
      \put(4663,484){\makebox(0,0){\strut{} 30}}%
      \put(5198,484){\makebox(0,0){\strut{} 35}}%
      \put(5733,484){\makebox(0,0){\strut{} 40}}%
      \put(6268,484){\makebox(0,0){\strut{} 45}}%
      \put(6803,484){\makebox(0,0){\strut{} 50}}%
      \put(22,2739){\rotatebox{-270}{\makebox(0,0){\strut{}$\delta$}}}%
      \put(4127,154){\makebox(0,0){\strut{}time step}}%
    }%
    \gplgaddtomacro\gplfronttext{%
    }%
    \gplbacktext
    \put(0,0){\includegraphics{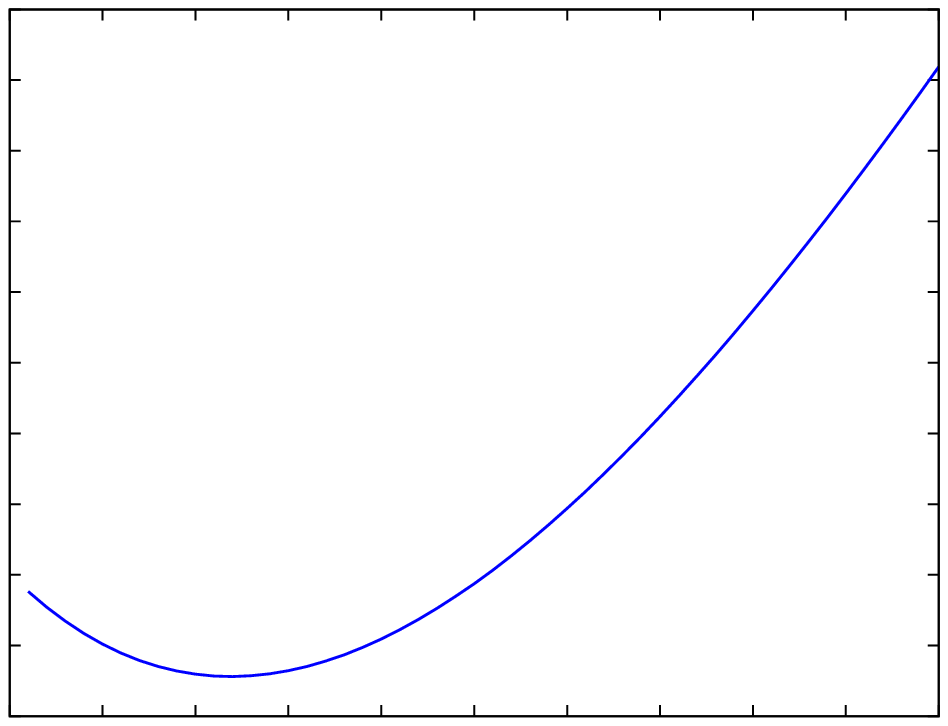}}%
    \gplfronttext
  \end{picture}%
\endgroup

%% file: redshift_test_cf_smooth.tex
\begingroup
  \makeatletter
  \providecommand\color[2][]{%
    \GenericError{(gnuplot) \space\space\space\@spaces}{%
      Package color not loaded in conjunction with
      terminal option `colourtext'%
    }{See the gnuplot documentation for explanation.%
    }{Either use 'blacktext' in gnuplot or load the package
      color.sty in LaTeX.}%
    \renewcommand\color[2][]{}%
  }%
  \providecommand\includegraphics[2][]{%
    \GenericError{(gnuplot) \space\space\space\@spaces}{%
      Package graphicx or graphics not loaded%
    }{See the gnuplot documentation for explanation.%
    }{The gnuplot epslatex terminal needs graphicx.sty or graphics.sty.}%
    \renewcommand\includegraphics[2][]{}%
  }%
  \providecommand\rotatebox[2]{#2}%
  \@ifundefined{ifGPcolor}{%
    \newif\ifGPcolor
    \GPcolortrue
  }{}%
  \@ifundefined{ifGPblacktext}{%
    \newif\ifGPblacktext
    \GPblacktexttrue
  }{}%
  \let\gplgaddtomacro\g@addto@macro
  \gdef\gplbacktext{}%
  \gdef\gplfronttext{}%
  \makeatother
  \ifGPblacktext
    \def\colorrgb#1{}%
    \def\colorgray#1{}%
  \else
    \ifGPcolor
      \def\colorrgb#1{\color[rgb]{#1}}%
      \def\colorgray#1{\color[gray]{#1}}%
      \expandafter\def\csname LTw\endcsname{\color{white}}%
      \expandafter\def\csname LTb\endcsname{\color{black}}%
      \expandafter\def\csname LTa\endcsname{\color{black}}%
      \expandafter\def\csname LT0\endcsname{\color[rgb]{1,0,0}}%
      \expandafter\def\csname LT1\endcsname{\color[rgb]{0,1,0}}%
      \expandafter\def\csname LT2\endcsname{\color[rgb]{0,0,1}}%
      \expandafter\def\csname LT3\endcsname{\color[rgb]{1,0,1}}%
      \expandafter\def\csname LT4\endcsname{\color[rgb]{0,1,1}}%
      \expandafter\def\csname LT5\endcsname{\color[rgb]{1,1,0}}%
      \expandafter\def\csname LT6\endcsname{\color[rgb]{0,0,0}}%
      \expandafter\def\csname LT7\endcsname{\color[rgb]{1,0.3,0}}%
      \expandafter\def\csname LT8\endcsname{\color[rgb]{0.5,0.5,0.5}}%
    \else
      \def\colorrgb#1{\color{black}}%
      \def\colorgray#1{\color[gray]{#1}}%
      \expandafter\def\csname LTw\endcsname{\color{white}}%
      \expandafter\def\csname LTb\endcsname{\color{black}}%
      \expandafter\def\csname LTa\endcsname{\color{black}}%
      \expandafter\def\csname LT0\endcsname{\color{black}}%
      \expandafter\def\csname LT1\endcsname{\color{black}}%
      \expandafter\def\csname LT2\endcsname{\color{black}}%
      \expandafter\def\csname LT3\endcsname{\color{black}}%
      \expandafter\def\csname LT4\endcsname{\color{black}}%
      \expandafter\def\csname LT5\endcsname{\color{black}}%
      \expandafter\def\csname LT6\endcsname{\color{black}}%
      \expandafter\def\csname LT7\endcsname{\color{black}}%
      \expandafter\def\csname LT8\endcsname{\color{black}}%
    \fi
  \fi
  \setlength{\unitlength}{0.0500bp}%
  \begin{picture}(7200.00,5040.00)%
    \gplgaddtomacro\gplbacktext{%
      \csname LTb\endcsname%
      \put(1320,704){\makebox(0,0)[r]{\strut{} 0.865}}%
      \put(1320,1286){\makebox(0,0)[r]{\strut{} 0.87}}%
      \put(1320,1867){\makebox(0,0)[r]{\strut{} 0.875}}%
      \put(1320,2449){\makebox(0,0)[r]{\strut{} 0.88}}%
      \put(1320,3030){\makebox(0,0)[r]{\strut{} 0.885}}%
      \put(1320,3612){\makebox(0,0)[r]{\strut{} 0.89}}%
      \put(1320,4193){\makebox(0,0)[r]{\strut{} 0.895}}%
      \put(1320,4775){\makebox(0,0)[r]{\strut{} 0.9}}%
      \put(1452,484){\makebox(0,0){\strut{} 10}}%
      \put(2047,484){\makebox(0,0){\strut{} 11}}%
      \put(2641,484){\makebox(0,0){\strut{} 12}}%
      \put(3236,484){\makebox(0,0){\strut{} 13}}%
      \put(3830,484){\makebox(0,0){\strut{} 14}}%
      \put(4425,484){\makebox(0,0){\strut{} 15}}%
      \put(5019,484){\makebox(0,0){\strut{} 16}}%
      \put(5614,484){\makebox(0,0){\strut{} 17}}%
      \put(6208,484){\makebox(0,0){\strut{} 18}}%
      \put(6803,484){\makebox(0,0){\strut{} 19}}%
      \put(286,2739){\rotatebox{-270}{\makebox(0,0){\strut{}$f$}}}%
      \put(4127,154){\makebox(0,0){\strut{}$\epsilon$ [MeV]}}%
    }%
    \gplgaddtomacro\gplfronttext{%
      \csname LTb\endcsname%
      \put(6212,4602){\makebox(0,0)[r]{\strut{}$f$ at $R_{\rm min}$}}%
      \csname LTb\endcsname%
      \put(6212,4382){\makebox(0,0)[r]{\strut{}$f$ at $R_{\rm max}$}}%
    }%
    \gplbacktext
    \put(0,0){\includegraphics{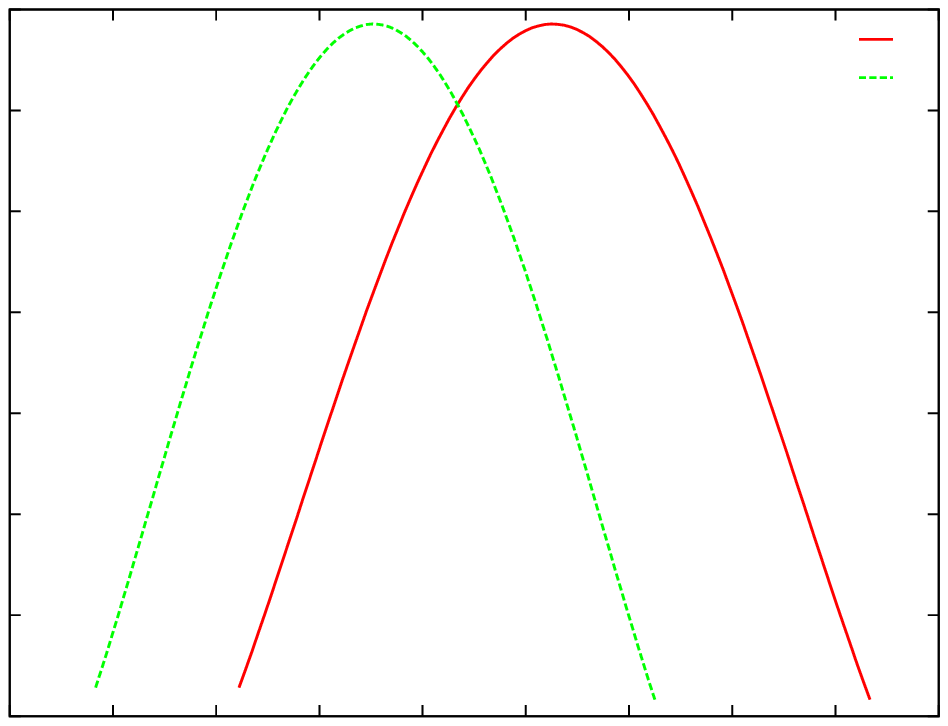}}%
    \gplfronttext
  \end{picture}%
\endgroup

%% file: searchlight.tex
\begingroup
  \makeatletter
  \providecommand\color[2][]{%
    \GenericError{(gnuplot) \space\space\space\@spaces}{%
      Package color not loaded in conjunction with
      terminal option `colourtext'%
    }{See the gnuplot documentation for explanation.%
    }{Either use 'blacktext' in gnuplot or load the package
      color.sty in LaTeX.}%
    \renewcommand\color[2][]{}%
  }%
  \providecommand\includegraphics[2][]{%
    \GenericError{(gnuplot) \space\space\space\@spaces}{%
      Package graphicx or graphics not loaded%
    }{See the gnuplot documentation for explanation.%
    }{The gnuplot epslatex terminal needs graphicx.sty or graphics.sty.}%
    \renewcommand\includegraphics[2][]{}%
  }%
  \providecommand\rotatebox[2]{#2}%
  \@ifundefined{ifGPcolor}{%
    \newif\ifGPcolor
    \GPcolortrue
  }{}%
  \@ifundefined{ifGPblacktext}{%
    \newif\ifGPblacktext
    \GPblacktexttrue
  }{}%
  \let\gplgaddtomacro\g@addto@macro
  \gdef\gplbacktext{}%
  \gdef\gplfronttext{}%
  \makeatother
  \ifGPblacktext
    \def\colorrgb#1{}%
    \def\colorgray#1{}%
  \else
    \ifGPcolor
      \def\colorrgb#1{\color[rgb]{#1}}%
      \def\colorgray#1{\color[gray]{#1}}%
      \expandafter\def\csname LTw\endcsname{\color{white}}%
      \expandafter\def\csname LTb\endcsname{\color{black}}%
      \expandafter\def\csname LTa\endcsname{\color{black}}%
      \expandafter\def\csname LT0\endcsname{\color[rgb]{1,0,0}}%
      \expandafter\def\csname LT1\endcsname{\color[rgb]{0,1,0}}%
      \expandafter\def\csname LT2\endcsname{\color[rgb]{0,0,1}}%
      \expandafter\def\csname LT3\endcsname{\color[rgb]{1,0,1}}%
      \expandafter\def\csname LT4\endcsname{\color[rgb]{0,1,1}}%
      \expandafter\def\csname LT5\endcsname{\color[rgb]{1,1,0}}%
      \expandafter\def\csname LT6\endcsname{\color[rgb]{0,0,0}}%
      \expandafter\def\csname LT7\endcsname{\color[rgb]{1,0.3,0}}%
      \expandafter\def\csname LT8\endcsname{\color[rgb]{0.5,0.5,0.5}}%
    \else
      \def\colorrgb#1{\color{black}}%
      \def\colorgray#1{\color[gray]{#1}}%
      \expandafter\def\csname LTw\endcsname{\color{white}}%
      \expandafter\def\csname LTb\endcsname{\color{black}}%
      \expandafter\def\csname LTa\endcsname{\color{black}}%
      \expandafter\def\csname LT0\endcsname{\color{black}}%
      \expandafter\def\csname LT1\endcsname{\color{black}}%
      \expandafter\def\csname LT2\endcsname{\color{black}}%
      \expandafter\def\csname LT3\endcsname{\color{black}}%
      \expandafter\def\csname LT4\endcsname{\color{black}}%
      \expandafter\def\csname LT5\endcsname{\color{black}}%
      \expandafter\def\csname LT6\endcsname{\color{black}}%
      \expandafter\def\csname LT7\endcsname{\color{black}}%
      \expandafter\def\csname LT8\endcsname{\color{black}}%
    \fi
  \fi
  \setlength{\unitlength}{0.0500bp}%
  \begin{picture}(7200.00,5040.00)%
    \gplgaddtomacro\gplbacktext{%
    }%
    \gplgaddtomacro\gplfronttext{%
      \csname LTb\endcsname%
      \put(1689,800){\makebox(0,0){\strut{} 0}}%
      \put(2573,800){\makebox(0,0){\strut{} 0.5}}%
      \put(3457,800){\makebox(0,0){\strut{} 1}}%
      \put(4339,800){\makebox(0,0){\strut{} 1.5}}%
      \put(5223,800){\makebox(0,0){\strut{} 2}}%
      \put(6107,800){\makebox(0,0){\strut{} 2.5}}%
      \put(3898,470){\makebox(0,0){\strut{}y [km]}}%
      \put(1518,1086){\makebox(0,0)[r]{\strut{}-3}}%
      \put(1518,1601){\makebox(0,0)[r]{\strut{}-2}}%
      \put(1518,2116){\makebox(0,0)[r]{\strut{}-1}}%
      \put(1518,2630){\makebox(0,0)[r]{\strut{} 0}}%
      \put(1518,3144){\makebox(0,0)[r]{\strut{} 1}}%
      \put(1518,3659){\makebox(0,0)[r]{\strut{} 2}}%
      \put(1518,4174){\makebox(0,0)[r]{\strut{} 3}}%
      \put(1188,2630){\rotatebox{-270}{\makebox(0,0){\strut{}z [km]}}}%
      \put(6570,1086){\makebox(0,0)[l]{\strut{}-0.02}}%
      \put(6570,1366){\makebox(0,0)[l]{\strut{}-0.01}}%
      \put(6570,1647){\makebox(0,0)[l]{\strut{} 0}}%
      \put(6570,1928){\makebox(0,0)[l]{\strut{} 0.01}}%
      \put(6570,2208){\makebox(0,0)[l]{\strut{} 0.02}}%
      \put(6570,2489){\makebox(0,0)[l]{\strut{} 0.03}}%
      \put(6570,2770){\makebox(0,0)[l]{\strut{} 0.04}}%
      \put(6570,3051){\makebox(0,0)[l]{\strut{} 0.05}}%
      \put(6570,3331){\makebox(0,0)[l]{\strut{} 0.06}}%
      \put(6570,3612){\makebox(0,0)[l]{\strut{} 0.07}}%
      \put(6570,3893){\makebox(0,0)[l]{\strut{} 0.08}}%
      \put(6570,4174){\makebox(0,0)[l]{\strut{} 0.09}}%
    }%
    \gplgaddtomacro\gplbacktext{%
    }%
    \gplgaddtomacro\gplfronttext{%
      \csname LTb\endcsname%
      \put(1689,800){\makebox(0,0){\strut{} 0}}%
      \put(2573,800){\makebox(0,0){\strut{} 0.5}}%
      \put(3457,800){\makebox(0,0){\strut{} 1}}%
      \put(4339,800){\makebox(0,0){\strut{} 1.5}}%
      \put(5223,800){\makebox(0,0){\strut{} 2}}%
      \put(6107,800){\makebox(0,0){\strut{} 2.5}}%
      \put(3898,470){\makebox(0,0){\strut{}y [km]}}%
      \put(1518,1086){\makebox(0,0)[r]{\strut{}-3}}%
      \put(1518,1601){\makebox(0,0)[r]{\strut{}-2}}%
      \put(1518,2116){\makebox(0,0)[r]{\strut{}-1}}%
      \put(1518,2630){\makebox(0,0)[r]{\strut{} 0}}%
      \put(1518,3144){\makebox(0,0)[r]{\strut{} 1}}%
      \put(1518,3659){\makebox(0,0)[r]{\strut{} 2}}%
      \put(1518,4174){\makebox(0,0)[r]{\strut{} 3}}%
      \put(1188,2630){\rotatebox{-270}{\makebox(0,0){\strut{}z [km]}}}%
      \put(6570,1086){\makebox(0,0)[l]{\strut{}-0.02}}%
      \put(6570,1366){\makebox(0,0)[l]{\strut{}-0.01}}%
      \put(6570,1647){\makebox(0,0)[l]{\strut{} 0}}%
      \put(6570,1928){\makebox(0,0)[l]{\strut{} 0.01}}%
      \put(6570,2208){\makebox(0,0)[l]{\strut{} 0.02}}%
      \put(6570,2489){\makebox(0,0)[l]{\strut{} 0.03}}%
      \put(6570,2770){\makebox(0,0)[l]{\strut{} 0.04}}%
      \put(6570,3051){\makebox(0,0)[l]{\strut{} 0.05}}%
      \put(6570,3331){\makebox(0,0)[l]{\strut{} 0.06}}%
      \put(6570,3612){\makebox(0,0)[l]{\strut{} 0.07}}%
      \put(6570,3893){\makebox(0,0)[l]{\strut{} 0.08}}%
      \put(6570,4174){\makebox(0,0)[l]{\strut{} 0.09}}%
    }%
    \gplbacktext
    \put(0,0){\includegraphics{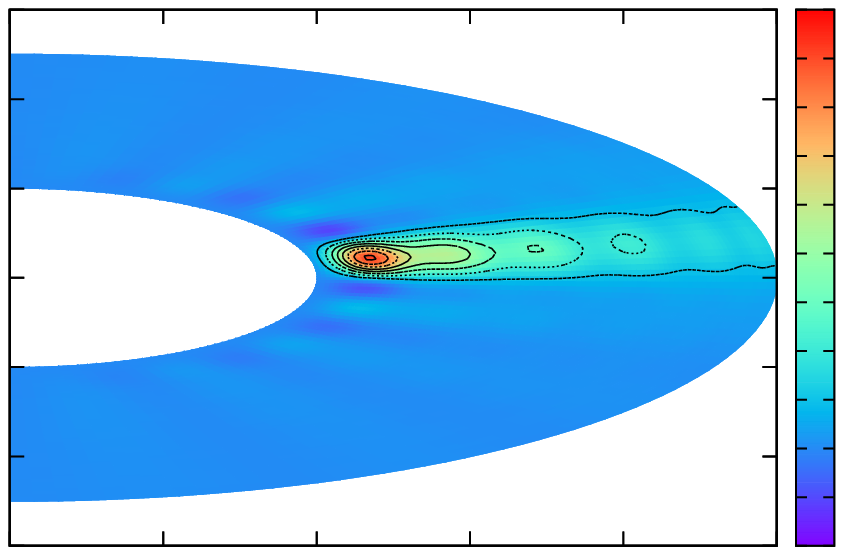}}%
    \gplfronttext
  \end{picture}%
\endgroup